\documentclass[preprint,12pt,authoryear]{elsarticle}

\usepackage{adjustbox}
\usepackage{booktabs}

\usepackage{adjustbox}   
\usepackage{float}       
\usepackage{placeins}    

\usepackage{amssymb}
\usepackage{booktabs}
\usepackage{amsmath}

\usepackage{adjustbox, siunitx, booktabs, amsmath}

\usepackage{booktabs}   
\usepackage{tabularx}   
\usepackage{makecell}   

\usepackage{booktabs} 
\usepackage{multirow} 
\usepackage{siunitx}  
\usepackage{array}
\usepackage{adjustbox}

\usepackage{multirow}
\usepackage[table,xcdraw]{xcolor}
\usepackage[most]{tcolorbox}
\usepackage{soul}
\definecolor{reddish}{HTML}{FBB4AE}
\definecolor{blueish}{HTML}{B3CDE3}
\definecolor{magentish}{HTML}{FF00AA}
\definecolor{greenish}{HTML}{a1d99b}
\definecolor{wpgreen}{HTML}{B5F7AC}
\definecolor{wppink}{HTML}{f8b1ee}
\definecolor{wpblue}{HTML}{c6c4fa}
\definecolor{wpyellow}{HTML}{fafab9}
\definecolor{wporange}{HTML}{fed6a6}

\begin{document}

\begin{frontmatter}



\title{Stress Bytes: Decoding the Associations between Internet Use and Perceived Stress} 


\author{Mohammad Belal, Nguyen Luong, Talayeh Aledavood, Juhi Kulshrestha} 

\affiliation[Mohammad Belal, Nguyen Luong, Talayeh Aledavood, Juhi Kulshrestha]{organization={Department of Computer Science Aalto University},
            addressline={Tietotekniikantalo, Konemiehentie 2}, 
            city={Espoo},
            postcode={ 02150}, 
            country={Finland}}

\begin{abstract}

In today’s digital era, internet plays a pervasive role in our lives, influencing everyday activities such as communication, work, and leisure. This online engagement intertwines with offline experiences, shaping individuals' overall well-being. Despite its significance, existing research often falls short in capturing the relationship between internet use and well-being, relying primarily on isolated studies and self-reported data. One of the major contributors to deteriorated well-being - both physical and mental – is stress. While some research has examined the relationship between internet use and stress, both positive and negative associations have been reported. Our primary goal in this work is to identify the associations between an individual’s internet use and their stress. For achieving our goal, we conducted a longitudinal multimodal study that spanned seven months. We combined fine-grained URL-level web browsing traces of 1490 German internet users with their sociodemographics and monthly measures of stress. Further, we developed a conceptual framework that allows us to simultaneously explore different contextual dimensions, including how, where, when, and by whom the internet is used. Our analysis revealed several associations between internet use and stress that vary by context. Social media, entertainment, online shopping, and gaming were positively associated with stress, while productivity, news, and adult content use were negatively associated. In the future, the behavioral markers we identified can pave the way for designing individualized tools for people to self-monitor and self-moderate their online behaviors to enhance their well-being, reducing the burden on already overburdened mental health services.

\end{abstract}



\begin{keyword}
Online behavior \sep Stress, \sep Internet use \sep Web browsing traces \sep Sociodemographic differences \sep Longitudinal design



\end{keyword}

\end{frontmatter}


\section{Introduction}
\label{Introduction}

Stress is an unavoidable part of human life, arising from the demands and challenges we face daily. It is a significant factor in health issues such as cardiovascular disease, weakened immune function, and mental health challenges \citep{Schneiderman2005,Bui2021-ht}. The internet, now an integral part of modern life, has sparked debates about its impact on stress levels and psychological well-being \citep{huang2010internet,ccikrikci2016effect}, as well as whether this influence is predominantly positive or negative. As our online and offline lives become increasingly interconnected, understanding the relationship between internet use and stress has gained considerable attention.

While the internet offers numerous advantages, such as enhanced connectivity and easy access to information, excessive or problematic use has been linked to various stress-inducing factors \citep{Jenaro2007, huang2010internet, Hkby2016}. On the other hand, past research suggests that not all aspects of internet use are detrimental; certain online activities, such as social communication and entertainment use, have been associated with reduced stress and improved psychological well-being \citep{ Nimrod2020, Luo2024,Heo2015}. 
Given the complex and multifaceted nature of stress and internet use, which have shown both positive and negative effects, a more comprehensive understanding of their relationship is needed. 


Despite the Internet's widespread influence, research on psychological well-being, including stress, has primarily focused on offline activities, leaving a critical gap in understanding how online behaviors impact stress and well-being \citep{Linton2016}. For instance, a comprehensive review of 99 commonly used psychological well-being scales identified 196 dimensions, yet none explicitly addressed online activities or behaviors \citep{Linton2016}. Moreover,  studies on online engagement have often faced limitations, including short study durations, small sample sizes, and an over-reliance on questionnaires to capture internet use patterns. These approaches can introduce biases and fail to provide a complete picture of the connection between online and offline experiences \citep{Kraut, Yetton2019-yj}.

Next, we will review previous research on the association between internet use and stress and examine the methodologies used in these studies. We will then outline our longitudinal multimodal study design, which integrates actual internet usage data with monthly questionnaires to measure stress, and discuss our study's potential impact.

\subsection{Conflicting Findings on Associations Between Stress and Internet Use}
 
The relationship between internet use and stress is complex, with prior research showing contrasting associations depending on the type and context of digital engagement, as well as individual characteristics. High levels of internet and smartphone use have been linked to increased stress \citep{Patel2019, nikolic2023smartphone}, often due to digital overload—the cognitive strain caused by constant notifications and an endless stream of information \citep{Reinecke2016, Barley2011}. In contrast, internet use through computers has been associated with less burnout compared to smartphone use \citep{Glda2022}. However, these effects are not uniform. For example, age moderates the impact of digital multitasking, with younger users reporting more stress than older adults when juggling multiple digital tasks \citep{Reinecke2016}. Other studies have shown no association \citep{Campbell2006}, or even a negative association between time spent online and stress, particularly in young adults \citep{stankovic2021}.

Moreover, the type of online activity plays a crucial role in stress outcomes. Social networking and entertainment-related use have been associated with higher stress levels, while internet use for work-related tasks has been linked to greater life satisfaction \citep{khalili2019each}. Research also indicates that communication overload from emails and messages is positively associated with perceived stress \citep{Reinecke2016}. Studies on social media show similarly nuanced findings. While Pew Research found no association between social media use and stress in men, a negative association was observed in women \citep{Hampton2015}. A large-scale study also showed slightly higher perceived stress among high social media users than non-users \citep{Nygaard2024}. Additionally, social media use has been found to delay recovery from real-world social stressors, suggesting a potential for stress maintenance or escalation \citep{Rus2017}.
Other digital behaviors, such as problematic news consumption \citep{McLaughlin2023} and adult content addiction \citep{grubbs2015internet}, have also been linked to heightened stress and emotional distress. Similarly, concerns such as cyber-bullying, online harassment \citep{Bezinovi2015}, work-life boundary erosion \citep{Bchler2020}, and data privacy \citep{Elhai2016} have been widely documented as stress-inducing. Further studies have identified links between stress and behaviors like online shopping addiction \citep{Li2022}, negative information seeking \citep{Kelly2024-1}, interpersonal communication \citep{Nimrod2020}, misinformation sharing \citep{verma2022examining}, and excessive gaming \citep{MUN2023107767, PARK2024108002}.

Conversely, the internet can also act as a buffer against stress \citep{stankovic2021}, offering access to supportive communities \citep{Heo2015}, relaxation tools, and leisure activities \citep{Nimrod2020}. Online entertainment and social interaction, in particular, have been shown to reduce stress and enhance well-being, especially among older populations \citep{Nimrod2020, Luo2024, rosell2022predictors}. Additionally, internet use has been recognized as a coping mechanism. Several online activities, including social media \citep{vanIngen2016}, entertainment \citep{Boursier2021, Nabi2017}, shopping \citep{Maraz2022}, and gaming \citep{vanIngen2016, pori2018}, have been identified in previous studies as strategies for managing stress.

\subsection{Methods for Identifying Connections between Internet Use and Stress}

Past research on internet use and stress has employed various methodologies. Many studies rely on cross-sectional designs and self-reported surveys \citep{Reinecke2016, Barley2011, khalili2019each, McLaughlin2023, Nimrod2020, MUN2023107767}. These studies often focus on specific populations, such as university or medical students \citep{Patel2019, nikolic2023smartphone}. Some have used larger samples \citep{McLaughlin2023, Nygaard2024, MUN2023107767} but still rely on questionnaires to capture internet use. A smaller number of experimental studies are available \citep{Rus2017, Kelly2024-1}, and some have adopted alternative approaches, such as analyzing social media data to infer psychological states \citep{verma2022examining}. Studies using actual web browsing data \citep{stankovic2021,yang2025studying} are limited and tend to capture general metrics, such as total time spent online \citep{stankovic2021}, and are based on relatively small samples (92 and 107 participants) indicating how difficult it is to conduct such studies.

\subsection{Contributions and Impact of Our Study}
\label{contributions}



Previous research on the relationship between internet use and stress has shown mixed findings, revealing both negative and positive associations depending on the type of internet activity. However, much of this evidence remains fragmented, as prior studies have largely relied on self-reported internet use data (which often lacks granularity) and have focused on limited aspects of internet use.

To address these limitations and to provide a more comprehensive understanding of how internet use relates to stress, we conduct a longitudinal multimodal study involving 1,490 internet users in Germany over seven months. Our study integrates fine-grained, passively collected web trace data from both desktop and mobile devices with participants’ monthly responses to a validated psychological stress scale (see Figure~\ref{study_design_figure}). Using objective behavioral data, we move beyond subjective self-reports and introduce a data-driven framework for revealing long-term usage patterns and identifying digital markers of stress.

Building on existing research, we further identify four key dimensions that shape the relationship between digital behavior and stress—
\begin{enumerate}
    \item \textbf{How}: The type and pattern of internet behaviors—such as usage volume \citep{nikolic2023smartphone, Thomee2010}, temporal rhythms \citep{Dissing2022}, and content categories (e.g., social media, productivity, entertainment, or shopping)—may influence stress in different ways \citep{Wolfers2022,Afifi2018,khalili2019each,Thomas2024}.
    
    \item \textbf{Where}: The device context plays a significant role. Prior research suggests that desktop use is often goal-oriented and structured, while mobile use tends to be more fragmented and reactive \citep{Oulasvirta2012}.
    
    \item \textbf{When}: The timing of online behaviors in relation to stress assessments is important, as short-term engagement with digital content may have immediate effects on stress responses \citep{Baryshnikov2023, Hu2016}.
    
    \item \textbf{By whom}: Individual differences—such as age, gender, income, and baseline stress levels—can moderate the impact of digital engagement on stress, with some groups being more susceptible than others \citep{Almeida2022Longitudinal,Johnson2023Perceived,Miquel2022The,Graves2021,Matud2004}.
\end{enumerate}

By adopting this multidimensional perspective, our study seeks to bring coherence to the scattered evidence in existing research and provide a more thorough understanding of the link between digital behaviors and stress. Additionally, identifying behavioral markers of stress in internet use may inform the design of future tools for real-time stress monitoring, complementing traditional self-reported measures. This approach contributes to a deeper understanding of digital well-being and supports the development of targeted interventions for healthier online behaviors.

\section{Methods}
\label{method}
In this section, we describe our study design (Section~\ref{study_section}) and participants (Section~\ref{participants}), the collected data (Section~\ref{data}) and extracted features (Section~\ref{measures}), and analysis models (Section~\ref{modeling}) that allow us to overcome the challenges of prior work described in previous section.

\subsection{Study Design}
\label{study_section}

\begin{figure}[htbp]
  \centering
  \resizebox{\textwidth}{!}{%
    \includegraphics{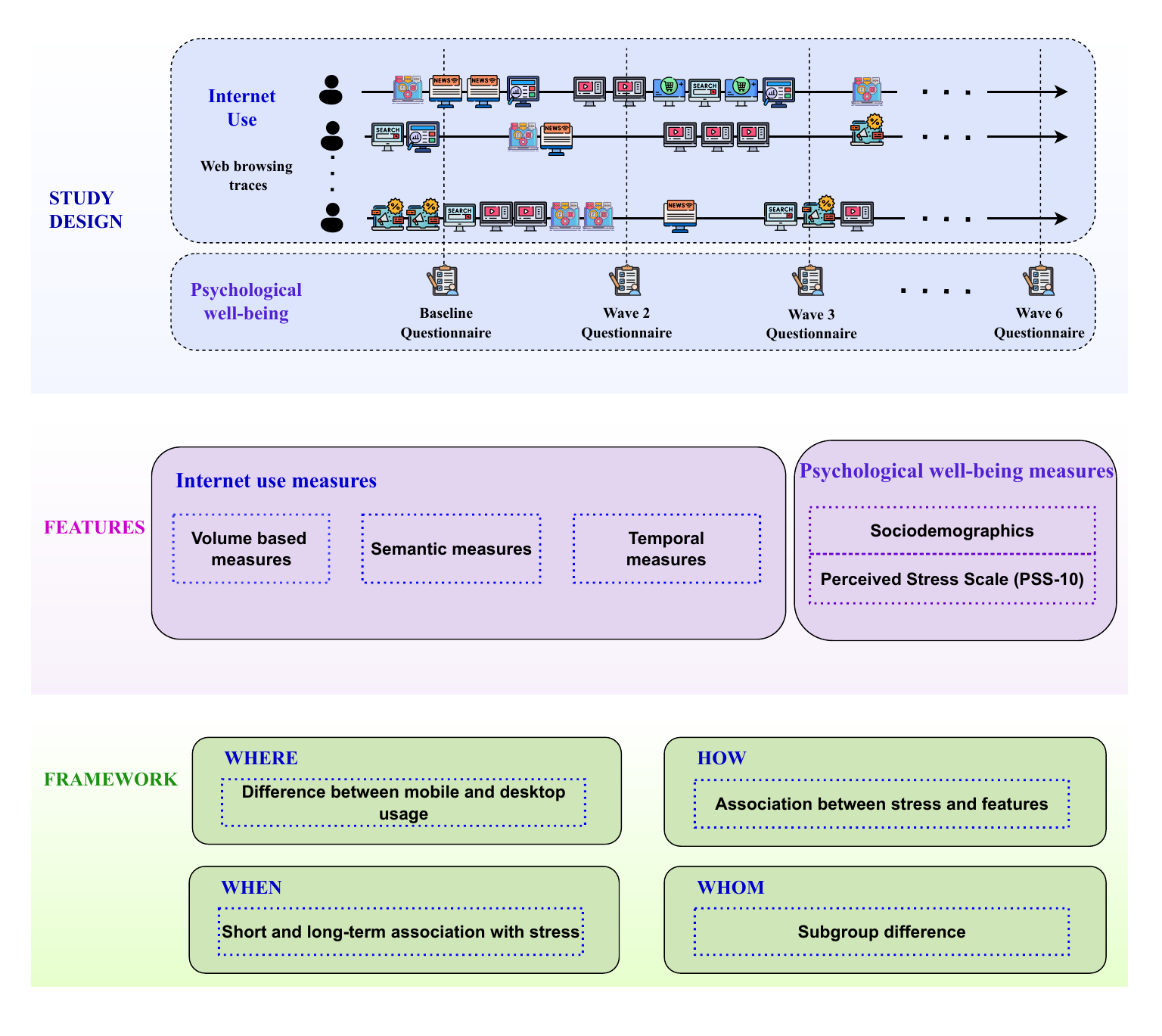}%
  }
  \caption{\textbf{Overview of the study design and contextual dimensions.} The top panel shows our longitudinal study design combining desktop and mobile web‐trace data with monthly stress questionnaires. The middle panel depicts the internet use and well-being features extracted from web browsing traces and monthly questionnaires. The bottom panel shows the contextual dimensions we consider for examining associations between internet use and stress.
}
\label{study_design_figure}
\end{figure}

We conducted a longitudinal multimodal study for seven months that combined passively collected fine-grained web browsing traces with repeated monthly online questionnaires (as depicted in Figure~\ref{study_design_figure}). The web browsing traces for desktop users included URL-level traces, while for mobile users, both URL-level and application-level traces were included.\footnote{We use app to denote mobile applications in the rest of the paper.} We measured the perceived stress of our panelists using the validated psychological Perceived stress scale (PSS-10) in six monthly waves. In the first wave, we also collected the sociodemographic characteristics of the panelists, including age, gender, and income. To ensure ethical compliance, we obtained ethics approval from our University's Research Ethics Committee (approval ID removed for double blind review) and got informed consent from the participants at each step. Moreover, we will share anonymized data and our code to support further research in the area.




\subsection{Participants}
\label{participants}

The study was conducted on a sample of German internet users, recruited through a GDPR-compliant panel company (Bilendi GmbH). All panelists from the company were invited for the first wave of questionnaires, and we got 1490 completed responses. In the subsequent five waves, all these 1490 respondents were invited to participate in each wave. Table~\ref{socio_demo} reports the number of participants with completed responses for each wave that we retained for further analysis.\footnote{We excluded some panelists from our analysis due to limited number of respondents in their category --one panelist that reported their gender to be non-binary, and 52 panelists that reported their income as `other' -- resulting in a total of 1437 respondents for the first wave.} We observe that about 23\% of the panelists drop out by the sixth wave. In Table~\ref{socio_demo}, we also depict the distribution of the panelists across age, gender and income for the six waves of questionnaires.

Next, we examined how closely our sample's sociodemographic distributions match German population margins for gender, age, and income (shown in Table~\ref{population_distribution}). We observe that our sample’s gender distribution matches closely with the German population’s. However, for both age and income, we observe that the middle ranges are overrepresented in our sample, while the extremes are underrepresented.




\begin{table*}[!htbp]
  \small
  \renewcommand{\arraystretch}{1.5}
  \centering
  \adjustbox{max width=\textwidth}{%
    \begin{tabular}{|l|c|c|c|c|c|c|}
      \hline
      \textbf{Wave}                    & \textbf{1}       & \textbf{2}       & \textbf{3}       & \textbf{4}       & \textbf{5}       & \textbf{6}       \\ \hline

      \textbf{Participants (n)}        & 1437             & 1314             & 1212             & 1198             & 1205             & 1107             \\ \hline

      \textbf{Gender}                  &                  &                  &                  &                  &                  &                  \\
      \hspace{5mm} Male (\%)           & 738 (51.35\%)    & 688 (52.36\%)    & 639 (52.72\%)    & 635 (53.01\%)    & 628 (52.12\%)    & 593 (53.57\%)    \\
      \hspace{5mm} Female (\%)         & 699 (48.65\%)    & 626 (47.64\%)    & 573 (47.28\%)    & 563 (46.99\%)    & 577 (47.88\%)    & 514 (46.43\%)    \\ \hline

      \textbf{Age Group (Years)}       &                  &                  &                  &                  &                  &                  \\
      \hspace{5mm} 18–30               & 119 (8.28\%)     & 101 (7.67\%)     & 91 (7.51\%)      & 84 (7.01\%)      & 92 (7.63\%)      & 76 (6.67\%)      \\
      \hspace{5mm} 31–45               & 462 (32.15\%)    & 414 (31.51\%)    & 385 (31.77\%)    & 382 (31.89\%)    & 379 (31.45\%)    & 330 (29.81\%)    \\
      \hspace{5mm} 46–60               & 569 (39.60\%)    & 528 (40.18\%)    & 483 (39.85\%)    & 483 (40.32\%)    & 483 (40.08\%)    & 460 (41.55\%)    \\
      \hspace{5mm} 60+                 & 287 (19.97\%)    & 271 (20.62\%)    & 253 (20.87\%)    & 249 (20.78\%)    & 251 (20.83\%)    & 241 (21.77\%)    \\ \hline

      \textbf{Income (Euros/month)}    &                  &                  &                  &                  &                  &                  \\
      \hspace{5mm} <1000 (Tier I)      & 126 (8.77\%)     & 119 (9.06\%)     & 116 (9.57\%)     & 110 (9.18\%)     & 103 (8.55\%)     & 94 (8.49\%)      \\
      \hspace{5mm} 1000–2000 (Tier II) & 300 (20.88\%)    & 271 (20.62\%)    & 244 (20.13\%)    & 244 (20.37\%)    & 249 (20.66\%)    & 240 (21.68\%)    \\
      \hspace{5mm} 2001–3000 (Tier III)& 364 (25.33\%)    & 339 (25.80\%)    & 319 (26.32\%)    & 310 (25.88\%)    & 309 (25.64\%)    & 281 (25.38\%)    \\
      \hspace{5mm} 3001–4000 (Tier IV) & 294 (20.46\%)    & 271 (20.62\%)    & 243 (20.05\%)    & 245 (20.45\%)    & 251 (20.83\%)    & 229 (20.69\%)    \\
      \hspace{5mm} 4000+ (Tier V)      & 353 (24.57\%)    & 314 (23.90\%)    & 290 (23.93\%)    & 289 (24.12\%)    & 293 (24.32\%)    & 263 (23.76\%)    \\ \hline

      \textbf{Perceived Stress Score (Mean (SD))} 
                                       & 16.19 (7.19)     & 15.83 (7.43)     & 15.76 (7.56)     & 15.65 (7.41)     & 15.54 (7.50)     & 14.89 (7.58)     \\ \hline
    \end{tabular}%
  }
  \caption{\textbf{Descriptive characteristics of participants across six survey waves.} The table presents the number of participants, gender distribution, age groups, income categories, and mean perceived stress scores (with standard deviations) for each wave. Percentages are shown in parentheses for categorical variables.}
  \label{socio_demo}
\end{table*}

\begin{table*}[!htbp]
  \footnotesize              
  \setlength{\tabcolsep}{3pt}
  \renewcommand{\arraystretch}{1.4}
  \centering
  \adjustbox{max width=0.4\textwidth}{
    \begin{tabular}{|l|c|}
      \hline
      \textbf{Adult Population}     & \textbf{Percentage (approx.)} \\ \hline
      Male                          & 48.8\%                        \\
      Female                        & 51.2\%                        \\ \hline
      \textbf{Age Group (Years)}    &                               \\ \hline
      18–30                         & 17\%                          \\
      31–45                         & 23\%                          \\
      46–60                         & 26\%                          \\
      60+                           & 34\%                          \\ \hline
      \textbf{Monthly Income Level} &                               \\ \hline
      Below €1250                   & 25.3\%                        \\
      €1250–€2080                   & 16.4\%                        \\
      €2080–€2920                   & 14.8\%                        \\
      €2920–€4170                   & 16.3\%                        \\
      Above €4170                   & 27.1\%                        \\ \hline
    \end{tabular}%
  }
  \caption{\textbf{Distribution of the adult population in Germany} (2023) by gender, age group, and monthly income level. Source: \citep{Destatis}}
  \label{population_distribution}
\end{table*}

\subsection{Data Collection}
\label{data}


The panelists of the panel company had already consented to install a tracking software on their desktops and/or mobile devices. Through this tracking software, the company provided fine-grained traces of visited URLs and mobile apps, including the time of visit and the duration of visit. During the seven-month period, we recorded 47,100,701 URL visits from both desktop and mobile users, covering 236,955 unique web domains. For mobile apps, we captured 13,553,645 app visits to 13,476 unique apps.




\begin{table*}[ht]
\renewcommand{\arraystretch}{1.5}
\centering
\resizebox{\textwidth}{!}{%
\begin{tabular}{|c|c|c|c|c|c|}
\hline
\textbf{Survey Wave} & \textbf{Number of Panelists} & \textbf{Desktop Users} & \textbf{Desktop Users (Cleaned)} & \textbf{Mobile Users} & \textbf{Mobile Users (Cleaned)} \\ \hline
1 & 1437 & 981 & 359 & 907 & 519 \\ \hline
2 & 1314 & 848 & 321 & 806 & 470 \\ \hline
3 & 1212 & 762 & 284 & 728 & 426 \\ \hline
4 & 1198 & 717 & 257 & 717 & 418 \\ \hline
5 & 1205 & 714 & 265 & 697 & 399 \\ \hline
6 & 1107 & 656 & 227 & 649 & 368 \\ \hline
\textbf{Total Distinct Users} & – & – & \textbf{526} & – & \textbf{656} \\ \hline
\end{tabular}%
}
\caption{\textbf{Overview of panelists with matched passive web data} from desktop and mobile devices across six survey waves. The table shows the number of users before and after data cleaning for both device types. The final row indicates the total number of distinct users retained in the cleaned dataset.}
\label{preprocessing}
\end{table*}

\begin{table*}[ht]
\renewcommand{\arraystretch}{1.5}
\centering
\resizebox{\textwidth}{!}{%
\begin{tabular}{|c|l|l|}
\hline
 & \textbf{Category} & \textbf{Example of Domains / Sub-domains and Apps in the Category} \\ \hline

\multirow{8}{*}{\textbf{Domains}} 
& Entertainment & youtube.com, twitch.tv, disneyplus.com, netflix.com \\ \cline{2-3}
& Shopping & amazon.de, ebay.de, kleinanzeigen.de, temu.com \\ \cline{2-3}
& Social Media & facebook.com, twitter.com, instagram.com \\ \cline{2-3}
& Messaging & whatsapp.com, knuddels.de, fdating.com \\ \cline{2-3}
& Productivity & mail.google.com, outlook.live.com, navigator.web.de, docs.google.com \\ \cline{2-3}
& Games & gameduell.de, anocris.com, forgeofempires.com, spielaffe.de \\ \cline{2-3}
& Adult & pornhub.com, xvideos.com, xnxx.com, romeo.com \\ \cline{2-3}
& News & bild.de, focus.de, welt.de, wunderweib.de \\ \hline

\multirow{7}{*}{\textbf{Apps}} 
& Entertainment & YouTube, Netflix, Spotify Music \\ \cline{2-3}
& Shopping & Amazon Shopping, eBay, Vinted.fr, Lidl Plus \\ \cline{2-3}
& Social Media & Facebook, Instagram, Twitter, TikTok – Make Your Day \\ \cline{2-3}
& Messaging & WhatsApp, Facebook Messenger, Telegram \\ \cline{2-3}
& Productivity & Gmail, GMX Mail, WEB.DE Mail, Google Calendar \\ \cline{2-3}
& Games & Candy Crush Saga, Coin Master, Royal Match, Pokémon GO \\ \cline{2-3}
& News & n-tv Nachrichten, kicker online, AOL – News, Apps \& Sites, BILD: Immer aktuell informiert \\ \hline

\end{tabular}%
}
\caption{\textbf{Categorization of web domains and mobile applications based on semantic usage type}. The table lists representative examples of domains and apps across various categories, grouped separately for desktop web domains and mobile applications.}
\label{semantic_category}
\end{table*}






\subsubsection{Data Cleaning \& Preprocessing}
\label{data_cleaning}
First, we removed the bottom 20\% of panelists in each wave ranked by total time spent browsing, since they did not have sufficient data to extract meaningful internet use patterns. Second, we observed that certain panelists were behaving as `professional survey-takers' with more than 25\% of their time online spent on survey domains. Therefore, we removed them from our sample to only retain users with more organic internet use. Lastly, to ensure that our internet use measures accurately capture user behavior, we only included panelists for whom we could categorize 80\% of their web visits (see Section~\ref{web_enrichment} for details).\footnote{We also tested that our results hold with lower thresholds (70\% and 75\%), confirming the robustness of our findings.} Table~\ref{preprocessing} summarizes the number of participants excluded at each step, resulting in a set of distinct panelists across waves comprising 656 mobile users and 526 desktop users.


\subsubsection{Data Enrichment}
\label{web_enrichment}
To understand \textit{how} the panelists are using the internet, we categorized their online visits into semantic categories. The goal was to group domains, sub-domains, and apps based on their primary function into categories such as `social media' (e.g., facebook.com, TikTok app) and `productivity' (e.g., Gmail, calendar.google.com). We derived the set of categories (listed in Table~\ref{semantic_category}) by combining categories used by app stores and web domain classification services such as webshrinker.com. For platform domains such as google.com, we also categorized their sub-domains. For instance, google.com was categorized as `search', while mail.google.com was classified as `productivity'. Table~\ref{semantic_category} presents the complete list of semantic categories we considered, along with some examples. 

Two researchers from our team first independently annotated the categories for all domains and apps that constituted around 85\% of web visits made by our panelists. Later, disagreements were resolved collaboratively. We observed a substantial inter-annotator agreement with a Cohen’s Kappa agreement \citep{Cohen1960} score of ~0.7, based on annotations for a random subset of 200 domains. Following this process, we classified 3,777 domains and 989 apps into semantic categories, capturing 85\% of visits from mobile devices and 84\% from desktops.

\subsection{Measures}
\label{measures}



\begin{table}
\centering
\resizebox{\textwidth}{!}{%
\begin{tabular}{|l|l|l|}
\hline
\textbf{} & \textbf{Description}                                                                                              & \textbf{Features}                                                                                                                                               \\ \hline
\textbf{Coarse-grained}           & Total time spent on online in period T                                                     & Total time spent online                                                                                                                                                    
 \\ 
 & &
 \\

& \begin{tabular}[c]{@{}l@{}} Difference of time spent online during daytime (06:00-18:00)\\ and nighttime hours(18:00-06:00)\end{tabular}                     & \begin{tabular}[c]{@{}l@{}}Daytime nighttime difference \end{tabular}                                                                       \\ \hline
\textbf{Fine-grained}         & \begin{tabular}[c]{@{}l@{}}Time spent on different semantic classes of online activities. \\For instance, time spent on entertainment domains or apps \\ like youtube.com or Amazon Prime is  classified as entertainment use.\end{tabular} & \begin{tabular}[c]{@{}l@{}}Time spent on entertainment\\ Time spent on social media\\ Time spent on messaging\\ Time spent on productivity\\ Time spent on shopping\\ Time spent on games \\Time spent on adult content \\ Time spent on news\end{tabular} \\ \hline
\textbf{Control variables}         & \begin{tabular}[c]{@{}l@{}}Sociodemographic characterisitics of individuals and seasonality \end{tabular}                     & \begin{tabular}[c]{@{}l@{}}Gender\\ Age\\ Income\\Survey wave\end{tabular}                                                                       \\ \hline

\end{tabular}%
}
\caption{\textbf{Features and their descriptions}. Time spent online is measured in hours in the period T (30 days or 2 days) before the measurement of stress. These features are computed for online activity on each device (desktop or mobile) separately.} 
\label{features_description}
\end{table}

As described in Section~\ref{study_section}, we combined repeated monthly psychological questionnaires with web browsing data. Perceived stress was measured through PSS-10 questionnaire responses, and internet use features were derived from passively collected web traces. For each panelist in each survey wave, we calculated internet usage features based on their activity during the period \(T\) preceding the stress measurement (i.e., the time of questionnaire response). To address the question of \textit{when} the internet is used, we extracted features for either 30 or 2 days, to examine both long-term and short-term effects. The resulting measures were then used to examine the associations between internet use and stress.


\subsubsection{Measures from Web Traces}

To measure \textit{how} individuals use the internet, we created features that span from coarse to fine-grained measures as shown in Table \ref{features_description}. We captured overall web activity at the coarse-grained level, such as total time spent online. We also accounted for the time of the day when panelists were browsing the web by including the difference between the time spent online during daytime (06:00-18:00) and nighttime (18:00-6:00) hours. At a finer granularity, we also examined how panelists divide their time between different semantic classes of online activities, such as engaging with social media or consuming entertainment or news.

\subsubsection{Measures from Questionnaires}

We use the Perceived Stress Scale-10 (PSS-10) \citep{Cohen1983} in our monthly questionnaires to measure the stress levels of our panelists. PSS-10 is a widely used, validated scale designed to assess how stressed individuals feel. It captures aspects such as the unpredictability of life, perceived control over situations, and general stress levels over the past month. Participants rate their responses on a scale from 0 (never) to 4 (very often), producing a total score between 0 and 40 across the 10 items. Higher scores indicate greater perceived stress, with scores typically grouped into three levels: 0–13 (low stress), 14–26 (moderate stress), and 27–40 (high stress)\citep{Philpott2022,Biswas2019}.
Additionally, we collected each participant's self-reported sociodemographics, including age, gender, and income, in the first wave of the questionnaires.



\subsection{Statistical Analysis}
\label{modeling}

We used Linear Mixed-Effects Models (LMM) ~\citep{Bryk2002} to examine the relationship between internet use and stress. We chose to use LMMs for analyzing the data from our longitudinal study since they account for repeated measurements of individuals and incorporate fixed and random effects. Fixed effects included internet use features described in Table ~\ref{features_description}. Random intercepts were added to account for individual-specific differences in baseline stress levels across participants.

We formally describe the models as follows -- 
for an individual  \( i \) at questionnaire wave \( j \) $\in$ \{1,2...6\}, we denote
\( Y_{ij} \) as the variable of interest, \( x_{ij} \) as the covariate, and the
intercept for the random effect as \( u_{j} \). Therefore, we consider the following LMM:

\[
Y_{ij} = \beta_0 + \beta_1 x_{ij1} + \beta_2 x_{ij2} + \dots + \beta_{n} x_{ijn} + u_j + \epsilon_{ij}
\]

where:
\begin{itemize}
    \item \( Y_{ij} \) is the perceived stress level of individual \( i \) measured via questionnaire wave \( j \).
    \item \( \beta_0 \) is the fixed intercept.
    \item \( \beta_1, \dots, \beta_n \) are the fixed effect coefficients for each covariate \( x_{ijn} \).
    \item \( x_{ij1} \) = total time spent online, \( x_{ij2} \) = daytime nighttime difference, \( x_{ij3} \) = time spent on entertainment ...  \( x_{ijn} \) =  survey wave, where \( x_{ijn} \) corresponds to each feature described in Table~\ref{features_description}.
    \item \( u_j \) is the random effect for individual \( i \), capturing individual-level variability.
    \item \( \epsilon_{ij} \) is the residual error term for individual \( i \) at wave \( j \).
\end{itemize}

We conducted model diagnostics to validate the assumptions of  LMMs, including checks for multicollinearity (vif $<$ 2.0). We implemented all statistical analyses using Python's statsmodels package (version 0.14.1) \citep{seabold2010statsmodels}.

To understand whether the granularity of the extracted features affects the model performance and the associations identified, we developed two models.
The first model ({\em Model 1}) focused on the coarse-grained measures of internet use such as total time spent online and daytime-nighttime difference. While the second model ({\em Model 2}) extended the first model and also incorporated finer-grained measures of internet use across semantic classes. For {\em Model 2}, we dropped the total time spent online feature to avoid multicollinearity.


Prior work has shown that both sociodemographics \citep{Almeida2022Longitudinal,Johnson2023Perceived,Miquel2022The,Graves2021,Matud2004} and seasonal variations \citep{Thorn2011, Gassen2024} can significantly influence individuals' stress levels. Accordingly, we included the sociodemographics and seasonality measures as control variables for both models, as detailed in Table~\ref{features_description}. 


\section{Results}

\begin{table}[htbp]
\centering
\begin{adjustbox}{max width=0.80\textwidth}
\begin{tabular}{l c r r c r r}
\toprule
\textbf{Predictors} & \multicolumn{3}{c}{Model 1} & \multicolumn{3}{c}{Model 2} \\
\cmidrule(lr){2-4} \cmidrule(lr){5-7}
& Estimate & CI & $p$ & Estimate & CI & $p$ \\
\midrule
Intercept & 20.77$^{***}$ & 19.09--22.44 & \textbf{$<$0.001} & 20.69$^{***}$ & 19.01--22.37 & \textbf{$<$0.001} \\
Survey wave & -0.10$^{*}$ & -0.19-- -0.02 & \textbf{0.014} & -0.11$^{*}$ & -0.19-- -0.02 & \textbf{0.011} \\
Gender & 1.79$^{***}$ & 0.80--2.77 & \textbf{$<$0.001} & 1.68$^{***}$ & 0.68--2.68 & \textbf{$<$0.001} \\
Age & -1.62$^{***}$ & -2.20-- -1.04 & \textbf{$<$0.001} & -1.57$^{***}$ & -2.16-- -0.99 & \textbf{$<$0.001} \\
Income & -1.11$^{***}$ & -1.50-- -0.73 & \textbf{$<$0.001} & -1.08$^{***}$ & -1.47-- -0.69 & \textbf{$<$0.001} \\
Total time spent online & 0.00 & -0.00--0.01 & 0.393 &  &  &  \\
Daytime nighttime difference & -0.01 & -0.02--0.00 & 0.124 & -0.01 & -0.02--0.00 & 0.175 \\
Time spent on entertainment &  &  &  & 0.01 & -0.01--0.02 & 0.365 \\
Time spent on social media &  &  &  & 0.00 & -0.01--0.02 & 0.710 \\
Time spent on messaging &  &  &  & 0.00 & -0.02--0.02 & 0.698 \\
Time spent on games &  &  &  & 0.00 & -0.00--0.01 & 0.296 \\
Time spent on shopping &  &  &  & 0.04$^{*}$ & 0.00--0.08 & \textbf{0.035} \\
Time spent on productivity &  &  &  & -0.03$^{*}$ & -0.06-- -0.00 & \textbf{0.042} \\
Time spent on news &  &  &  & -0.03 & -0.09--0.03 & 0.333 \\
\midrule
\multicolumn{7}{l}{\textbf{Random Effects}} \\
$\sigma^2$ & \multicolumn{3}{c}{11.67} & \multicolumn{3}{c}{11.65} \\
$\tau_{00}$ & \multicolumn{3}{c}{36.54$_{\text{pid}}$} & \multicolumn{3}{c}{36.48$_{\text{pid}}$} \\
ICC & \multicolumn{3}{c}{0.76} & \multicolumn{3}{c}{0.76} \\
N & \multicolumn{3}{c}{656$_{\text{pid}}$} & \multicolumn{3}{c}{656$_{\text{pid}}$} \\
Observations & \multicolumn{3}{c}{2600} & \multicolumn{3}{c}{2600} \\
\bottomrule
                \multicolumn{7}{l}{\textit{\textbf{* $p<0.05$ \quad ** $p<0.01$ \quad *** $p<0.001$}}} \\
                \end{tabular}
                \end{adjustbox}
                
\caption{\textbf{Results from linear mixed-effects models for all participants, based on 30-days mobile data.} Model 1 includes coarse-grained features, while Model 2 incorporates fine-grained usage categories (described in Section \ref{modeling}). Estimates, confidence intervals (CI), and p-values are reported for each predictor. Statistically significant p-values are in bold. Random effects, intraclass correlation coefficient (ICC), number of participants (N), and total observations are also provided.}
\label{tab:mixed_effects_all_participants for last 30 of mobile data}
                \end{table}

\begin{table}[htbp]
\centering
\begin{adjustbox}{max width=0.80\textwidth}
\begin{tabular}{l c r r c r r}

\toprule
\textbf{Predictors} & \multicolumn{3}{c}{Model 1} & \multicolumn{3}{c}{Model 2} \\
\cmidrule(lr){2-4} \cmidrule(lr){5-7}
& Estimate & CI & $p$ & Estimate & CI & $p$ \\
\midrule
Intercept & 20.62$^{***}$ & 18.70--22.55 & \textbf{$<$0.001} & 20.49$^{***}$ & 18.58--22.41 & \textbf{$<$0.001} \\
Survey wave & -0.14$^{**}$ & -0.24-- -0.03 & \textbf{0.009} & -0.15$^{**}$ & -0.26-- -0.04 & \textbf{0.006} \\
Gender & 1.83$^{**}$ & 0.66--3.00 & \textbf{0.002} & 1.73$^{**}$ & 0.55--2.90 & \textbf{0.004} \\
Age & -1.72$^{***}$ & -2.41-- -1.04 & \textbf{$<$0.001} & -1.74$^{***}$ & -2.43-- -1.05 & \textbf{$<$0.001} \\
Income & -0.94$^{***}$ & -1.38-- -0.51 & \textbf{$<$0.001} & -0.92$^{***}$ & -1.36-- -0.48 & \textbf{$<$0.001} \\
Total time spent online & -0.00 & -0.01--0.00 & 0.408 &  &  &  \\
Daytime nighttime difference & 0.00 & -0.01--0.01 & 0.461 & 0.00 & -0.01--0.01 & 0.459 \\
Time spent on entertainment &  &  &  & -0.00 & -0.01--0.01 & 0.640 \\
Time spent on adult content &  &  &  & -0.01 & -0.03--0.00 & 0.108 \\
Time spent on social media &  &  &  & 0.00 & -0.02--0.02 & 0.937 \\
Time spent on messaging &  &  &  & 0.01 & -0.04--0.06 & 0.608 \\
Time spent on games &  &  &  & 0.00 & -0.03--0.03 & 0.836 \\
Time spent on shopping &  &  &  & 0.03 & -0.00--0.06 & 0.090 \\
Time spent on productivity &  &  &  & -0.00 & -0.03--0.02 & 0.855 \\
Time spent on news &  &  &  & -0.02 & -0.06--0.02 & 0.283 \\
\midrule
\multicolumn{7}{l}{\textbf{Random Effects}} \\
$\sigma^2$ & \multicolumn{3}{c}{11.10} & \multicolumn{3}{c}{11.09} \\
$\tau_{00}$ & \multicolumn{3}{c}{40.62$_{\text{pid}}$} & \multicolumn{3}{c}{40.77$_{\text{pid}}$} \\
ICC & \multicolumn{3}{c}{0.79} & \multicolumn{3}{c}{0.79} \\
N & \multicolumn{3}{c}{526$_{\text{pid}}$} & \multicolumn{3}{c}{526$_{\text{pid}}$} \\
Observations & \multicolumn{3}{c}{1713} & \multicolumn{3}{c}{1713} \\
\bottomrule
                \multicolumn{7}{l}{\textit{\textbf{* $p<0.05$ \quad ** $p<0.01$ \quad *** $p<0.001$}}} \\
                \end{tabular}
                \end{adjustbox}
                
\caption{\textbf{Results from linear mixed-effects models for all participants, based on 30-days desktop data.} Model 1 includes coarse-grained features, while Model 2 incorporates fine-grained usage categories (described in Section \ref{modeling}). Estimates, confidence intervals (CI), and p-values are reported for each predictor. Statistically significant p-values are in bold. Random effects, intraclass correlation coefficient (ICC), number of participants (N), and total observations are also provided.}
\label{tab:mixed_effects_all_participants for last 30 of desktop data}
                \end{table}

Our study examined various internet-use behaviors associated with stress, and in this section, we present our results across four key contextual dimensions (as outlined in Section~\ref{contributions}). We began by first analyzing \textit{how} internet-based features relate to stress. We then explored the remaining dimensions: device-based differences (\textit{where}) by comparing desktop and mobile usage, temporal patterns (\textit{when}) using internet activity from the two days prior to the survey, and individual differences (\textit{by whom}) through subgroup analyses based on age, gender, income, and baseline stress levels.

\subsection{Behavioral Patterns: (How)}

To understand how internet usage is associated with stress, we ran linear mixed-effects models on two sets of features, progressing from coarse-grained (amount and timing of usage) to fine-grained (also including semantic category usage) measures. An ANOVA test was conducted to determine whether the more complex model explained significantly more variance than the simpler model. The results showed no statistically significant improvement when using the more complex model, but the use of more complex model provided important information into the nuanced relationship between internet use and stress.

Analysis of 30-day mobile data (Number of Panelists \( N = 656 \), Observations \( n = 2600 \)), as shown in Table~\ref{tab:mixed_effects_all_participants for last 30 of mobile data}, revealed that Model 2—which includes both timing and semantic category usage—identified significant associations with stress. Specifically, shopping-related usage was positively associated with stress (\( \beta = 0.04 \), CI = [0.00–0.08], \( p = 0.035 \)), while productivity usage showed a negative association (\( \beta = -0.03 \), CI = [-0.06– -0.00], \( p = 0.042 \)). In contrast, Model 1, which included only total usage and timing, did not show any significant associations.

Additionally, socio-demographic factors such as age, gender, and income consistently predicted stress across both models. Age (\( \beta = -1.57 \),CI = [-2.16– -0.99], \( p < 0.001 \)) and income (\( \beta = -1.08 \), CI = [-1.47– -0.69], \( p < 0.001 \)) were negatively associated with stress, while women reported higher stress levels (\( \beta = 1.68 \),CI = [0.68–2.68] \( p = 0.001 \)).

\subsection{Device Matters (Where)}
To observe device differences, we analyzed 30-day desktop data. For desktop data (\( N = 526 \), \( n = 1713 \)), the results are shown in Table \ref{tab:mixed_effects_all_participants for last 30 of desktop data}. Model 2, which incorporates semantic and temporal features, showed a weaker positive association between shopping usage and stress (\( \beta = 0.03 \), CI = [-0.0–0.06] \( p = 0.090 \)). As observed with mobile data, the simpler Model 1 did not reveal any significant associations with internet usage features. Similarly, the socio-demographics results were consistent with those observed with mobile data.

\subsection{Time Period of Data (When)}

To investigate the relationship between short-term vs. long-term internet usage patterns and stress, we analyzed associations between various online activities performed on mobile and desktop devices in two time periods -- 30 days and 2 days -- and individual stress levels. We employed the same features and models as in our previous 30 day data analyses. Here, we, specifically, focused on web activity recorded during the two days immediately preceding the PSS-10 survey.

For both mobile and desktop data (as shown in Tables \ref{tab:mixed_effects_all_participants for last 2 of mobile data} and \ref{tab:mixed_effects_all_participants for last 2 of desktop data} in the Appendix), news usage showed a negative association with stress (\( \beta = -0.54 \),CI = [-1.08–0.00] \( p = 0.048 \)) and (\( \beta = -0.50 \),CI = [-0.90– -0.11] \( p = 0.012 \)) respectively in model 2. Additionally, in desktop data, messaging usage demonstrated a weak negative association with stress (\( \beta = -0.59 \), CI = [-1.24–0.06] \( p = 0.073 \)) in model 2.

\subsection{Individual Differences (By Whom)}
To explore how internet usage varies by individual characteristics, we conducted subgroup analyses, running models separately for categories such as gender (male and female) to understand how associations differ based on these characteristics. In the following subsections, we first examined the relationship between internet use and stress based on baseline stress levels by analyzing high-stress and low-stress groups. We then explored differences by gender, age, and income categories.

\subsubsection{Stress Levels}
We identified two groups of panelists from our data -- high-stress and low-stress -- based on their reported PSS-10 scores in the online questionnaires. The panelists who scored more than 26 in each wave they participated in were included in the high-stress group. And the panelists who scored below 14 in each wave they participated in were included in the low-stress group.

For the high-stress population (\( \text{PSS-10 score} > 26 \), several notable results were observed (as shown in Tables \ref{tab:mixed_effects_high_stress for last 30 of mobile data} -- \ref{tab:mixed_effects_high_stress for last 2 of desktop data} in the Appendix). In the 30-day mobile data, time spent (\( \beta = 0.01 \), CI = [0.0–0.02], \( p < 0.001 \)) in model 1, and social media usage (\( \beta = 0.02 \), CI = [0.0–0.03], \( p = 0.021 \)), and gaming usage (\( \beta = 0.01 \), CI = [0.0–0.02], \( p = 0.021 \)) in model 2 were positively associated with stress. In the two days of data, the daytime-nighttime difference showed a weak positive association (\( \beta = 0.11 \), CI = [-0.01–0.23], \( p = 0.075 \)) in model 1. For desktop data, no significant variables, including socio-demographics, were found to be associated with stress in the high-stress subgroup.

In the low-stress population (\( \text{PSS-10 score} < 14 \) in the baseline survey) as shown in Tables \ref{tab:mixed_effects_low_stress for last 30 of mobile data} --\ref{tab:mixed_effects_low_stress for last 2 of desktop data} in the Appendix, adult content usage was negatively associated with stress in 30-day desktop data (\( \beta = -0.02 \), CI = [-0.04–0.0], \( p = 0.065 \)). Similarly, for the low-stress group, in the two days data, time spent (\( \beta = -0.07 \), CI = [-0.14–0.0], \( p = 0.038 \)) was significant for desktop, while gaming usage was weakly significant for mobile data (\( \beta = 0.08 \), CI = [-0.01–0.17], \( p = 0.098 \)).

When analyzing the 30-days data for all participants, socio-demographic factors were strongly associated with stress in both mobile and desktop settings. However, within the high-stress group, income was the only socio-demographic variable that remained significant in mobile data, showing a negative association with stress in both the general population (\( \beta = -1.08 \), CI = [-1.47– -0.69], \( p < 0.001 \)) and the high-stress subgroup (\( \beta = -0.52 \), CI = [-0.91– -0.12], \( p = 0.010 \)). In contrast, gender and age, which were significant predictors in the overall population, did not show statistical significance in the highly stressed subgroup for either mobile or desktop data.

\subsubsection{Gender Differences}

Subgroup analysis by gender revealed distinct patterns in feature significance for both desktop and mobile data. In the 30-day mobile data (as shown in Table \ref{tab:mixed_effects_male for last 30 of mobile data} in the Appendix), shopping usage (\( \beta = 0.07 \), CI = [0.01– 0.14], \( p = 0.022 \)) and productivity features (\( \beta = -0.05 \), CI = [-0.09– -0.01], \( p = 0.021 \)) were significant only for male users (\( N = 334 \) panelists, \( n = 1,334 \) observations) in model 2. In contrast, these features were not significant for female users (\( N = 322 \) panelists, \( n = 1,266 \) observations) as shown in Table \ref{tab:mixed_effects_female for last 30 of mobile data} in the Appendix. No features were significant for males and females in the 30-day desktop data.

In the 2-day data (as shown in Tables \ref{tab:mixed_effects_male for last 2 of mobile data}--\ref{tab:mixed_effects_female for last 2 of mobile data} in the Appendix), news consumption was negatively associated with stress for males in both desktop (\( \beta = -0.52 \), CI = [-1.02– -0.01], \( p = 0.044 \)) and mobile (\( \beta = -0.58 \), CI = [-1.23– 0.07], \( p = 0.079 \)) data. For females, in mobile data, daytime-nighttime difference (\( \beta = 0.1 \), CI = [-0.0–0.21], \( p = 0.060 \)) and messaging (\( \beta = 0.23 \), CI = [-0.02– 0.48], \( p = 0.075 \)) were weakly positively associated.

\subsubsection{Age Differences}

Subgroup analysis by age revealed distinct patterns in web-based associations. For the 30-day mobile data (as shown in Tables \ref{tab:mixed_effects_age_18–30 for last 30 of mobile data}-- \ref{tab:mixed_effects_age_60+ for last 30 of mobile data} in the Appendix), shopping was positively associated with stress in the 18-30 (\( \beta = 0.12 \), CI = [-0.02–0.25], \( p = 0.085 \)) and 60+ (\( \beta = 0.10 \), CI = [0.02–0.19], \( p = 0.021 \)) age groups. In the 30-45 age group, weak positive associations were found for entertainment (\( \beta = 0.02 \), CI = [-0.00–0.04], \( p = 0.075 \)) and messaging (\( \beta = 0.03 \), CI = [-0.00–0.06], \( p = 0.066 \)), whereas productivity was negatively associated (\( \beta = -0.08 \), CI = [-0.14– -0.02], \( p = 0.007 \)). In the 60+ age group, time spent (\( \beta = 0.01 \), CI = [0.00–0.03], \( p = 0.029 \)) and messaging (\( \beta = 0.05 \), CI = [-0.00–0.09], \( p = 0.057 \)) were positively associated, while daytime-nighttime difference (\( \beta = -0.03 \), CI = [-0.05– -0.01], \( p = 0.012 \)) and shopping (\( \beta = 0.10 \), CI = [0.02–0.19], \( p = 0.021 \)) were negatively associated.

In the 30-day desktop data (as shown in Tables \ref{tab:mixed_effects_age_18–30 for last 30 of desktop data}-- \ref{tab:mixed_effects_age_60+ for last 30 of desktop data} in the Appendix), adult content usage was negatively associated in the 18-30 (\( \beta = -0.87 \), CI = [-1.77–0.03], \( p = 0.058 \)) and 60+ (\( \beta = -0.19 \), CI = [-0.38–0.00], \( p = 0.055 \)) age groups. Additionally, in the 18-30 age group, daytime-nighttime difference (\( \beta = 0.07 \), CI = [-0.00–0.14], \( p = 0.058 \)) was positively associated, while news usage (\( \beta = -0.26 \), CI = [-0.42– -0.10], \( p = 0.002 \)) was negatively associated. Shopping was positively associated in the 45-60 age group (\( \beta = 0.04 \), CI = [-0.00–0.09], \( p = 0.062 \)).

For the 2-day mobile data (as shown in Tables \ref{tab:mixed_effects_age_18–30 for last 2 of mobile data}-- \ref{tab:mixed_effects_age_60+ for last 2 of mobile data} in the Appendix), news was negatively associated in both the 30-45 (\( \beta = -1.01 \), CI = [-2.22– 0.19], \( p = 0.098 \)) and 45-60 (\( \beta = -0.85 \), CI = [-1.76– 0.06], \( p = 0.068 \)) age groups. Additionally, in the 30-45 age group, daytime-nighttime difference (\( \beta = 0.14 \), CI = [0.02– 0.25], \( p = 0.020 \)) and entertainment usage (\( \beta = 0.22 \), CI = [0.01– 0.43], \( p = 0.039 \)) were positively associated, whereas shopping was negatively associated (\( \beta = -0.54 \), CI = [-1.08– -0.0], \( p = 0.050 \)). In 60+ age group time spent on gaming (\( \beta = 0.26 \), CI = [0.01– 0.50], \( p = 0.043 \)) was positively associated.

Similarly, for the 2-day desktop data (as shown in Tables \ref{tab:mixed_effects_age_18–30 for last 2 of desktop data}-- \ref{tab:mixed_effects_age_60+ for last 2 of desktop data} in the Appendix), messaging was negatively associated in the 45-60 age group (\( \beta = -0.78 \), CI = [-1.48– -0.07], \( p = 0.031 \)), but positively associated for the 60+ age group (\( \beta = 7.63 \), CI = [0.39–14.87], \( p = 0.039 \)). In the 18-30 age group, entertainment (\( \beta = 1.14 \), CI = [-0.18– 2.46], \( p = 0.090 \)) was positively associated, and social media (\( \beta = 1.22 \), CI = [0.15–2.29], \( p = 0.025 \)) was positive for the 30-45 age group. Additionally, time spent (\( \beta = -0.14 \), CI = [0.28– 0.0], \( p = 0.058 \)) and news usage (\( \beta = 0.46 \), CI = [-0.96– 0.04], \( p = 0.071 \)) were negatively associated in the 60+ age group.



\subsubsection{Income Differences}

For the 30-day mobile data (as shown in Tables \ref{tab:Results from mixed-effects models for tier 1, based on 30 days of mobile usage data.}-- \ref{tab:Results from mixed-effects models for tier 5, based on 30 days of mobile usage data.} in the Appendix), messaging (\( \beta = -0.06 \), CI = [-0.12– -0.01], \( p = 0.030 \)) was negatively associated with stress in participants earning less than €1,000 per month (Tier I). In the €2,001–€3,000 income group (Tier II), productivity (\( \beta = -0.05 \), CI = [-0.10– 0.0], \( p = 0.059 \)) and news usage (\( \beta = -0.10 \), CI = [-0.20– 0.01], \( p = 0.078 \)) were both negatively associated. Time spent (\( \beta = 0.01 \), CI = [0.0– 0.02], \( p = 0.015 \)) and shopping (\( \beta = 0.08 \), CI = [0.01– 0.16], \( p = 0.022 \)) were positively associated with stress for participants earning €3,001–€4,000 (Tier IV). No other significant internet-based features were observed for other income categories.

For the 30-day desktop data (as shown in Tables \ref{tab:Results from mixed-effects models for tier 1, based on 30 days of desktop usage data.}--\ref{tab:Results from mixed-effects models for tier 5, based on 30 days of desktop usage data.} in the Appendix), news usage was positively associated with stress in Tier I income group  (\( \beta = 0.28 \), CI = [0.05– 0.50], \( p = 0.015 \)), while productivity was negatively associated (\( \beta = -0.07 \), CI = [-0.13– -0.01], \( p = 0.030 \)). For participants in Tier III income, news usage (\( \beta = -0.13 \), CI = [-0.25– 0.0], \( p = 0.056 \)) was negatively associated. For participants in Tier IV income, shopping was positively associated with stress (\( \beta = 0.08 \), CI = [0.02– 0.15], \( p = 0.016 \)). For Tier V participants, social media use was positively associated with stress (\( \beta = 0.06 \), CI = [0.01– 0.12], \( p = 0.026 \)), while news use  (\( \beta = -0.16 \), CI = [-0.25– -0.07], \( p < 0.001 \)) and time spent were negatively associated (\( \beta = -0.01 \), CI = [-0.03– 0.0], \( p = 0.025 \)). No significant associations were identified for other income categories.

For the 2-day mobile data (as shown in Tables \ref{tab:Results from mixed-effects models for tier 1, based on 2 days of mobile usage data.}--\ref{tab:mixed_effects_Tier V for last 2 of mobile data} in the Appendix), gaming (\( \beta = 0.21 \), CI = [-0.02– 0.43], \( p = 0.072 \)) was positively associated in the Tier II income group and negatively associated (\( \beta = -0.19 \), CI = [-0.37– 0.0], \( p = 0.047 \)) in the Tier V income group. News use was negatively associated in the Tier III (\( \beta = -0.75 \), CI = [-1.58– 0.09], \( p = 0.081 \)) and Tier V (\( \beta = -0.74 \), CI = [-1.60– 0.12], \( p = 0.090 \)) income groups. Additionally, messaging (\( \beta = 0.48 \), CI = [0.14– 0.82], \( p = 0.006 \)) was positively associated in the Tier III income group, and daytime-nighttime difference (\( \beta = 0.18 \), CI = [0.0– 0.36], \( p = 0.044 \)) was positively associated in the Tier IV income group.

For the 2-day desktop data (as shown in Tables \ref{tab:Results from mixed-effects models for tier 1, based on 2 days of desktop usage data.}-\ref{tab:Results from mixed-effects models for tier 5, based on 2 days of desktop usage data.} in the Appendix), news usage was negatively associated in the Tier III income group (\( \beta = -1.13 \), CI = [-2.33–0.06], \( p = 0.063 \)) and Tier V income group (\( \beta = -1.04 \), CI = [-1.76– -0.32], \( p = 0.005 \)). Additionally, in the Tier V income group, productivity (\( \beta = 0.64 \), CI = [0.07–1.21], \( p = 0.027 \)) was positively associated with stress. In the Tier IV income group, time spent (\( \beta = 0.23 \), CI = [0.04–0.42], \( p = 0.019 \)) and daytime-nighttime difference (\( \beta = 0.23 \), CI = [-0.00– 0.46], \( p = 0.050 \)) were positively associated.

\section{Discussion}
\begin{figure}[htbp]
  \centering
  \resizebox{\textwidth}{!}{%
    \includegraphics{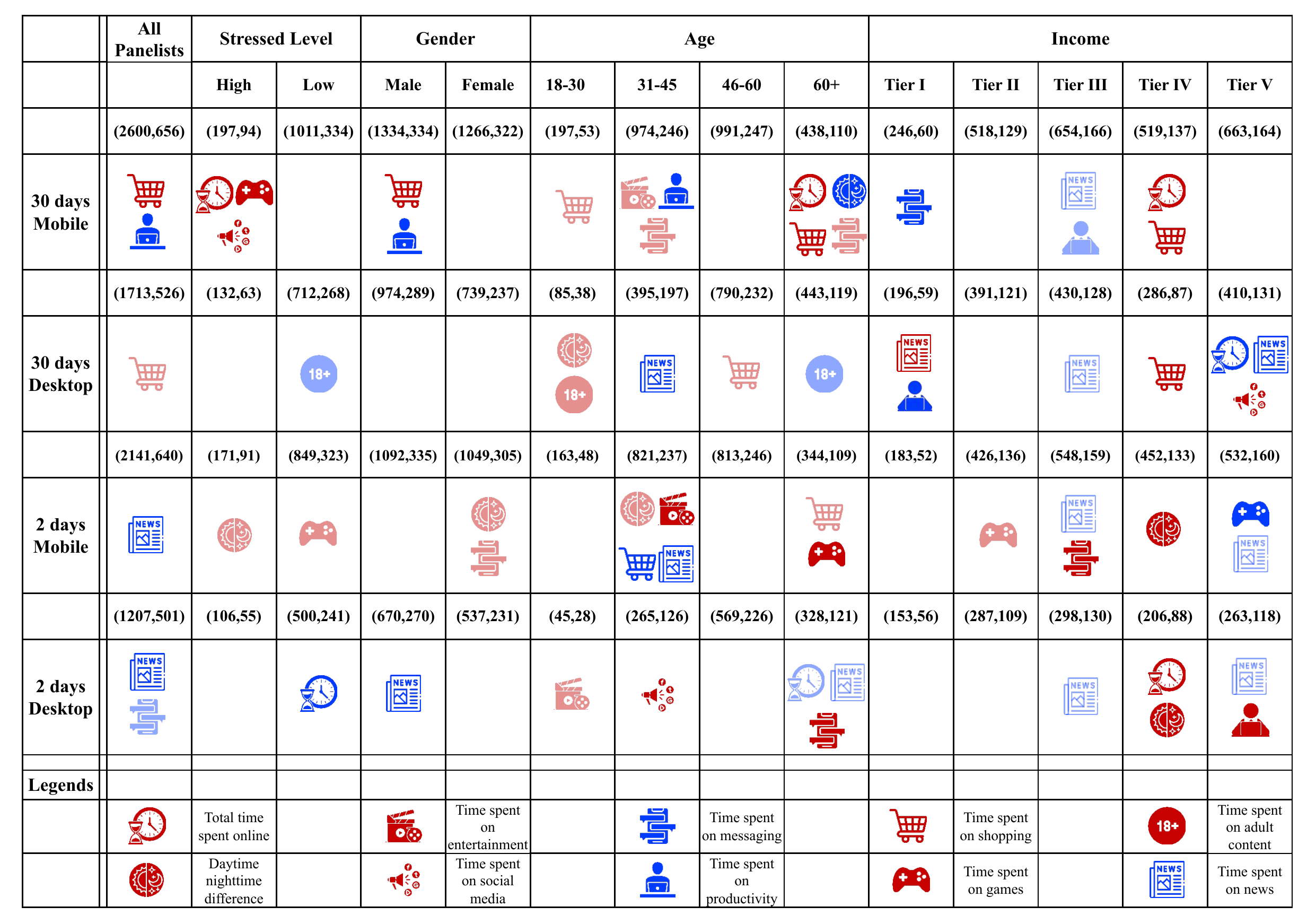}%
  }
  \caption{\textbf{Overview of significant internet usage behaviors across various contextual dimensions.} The icons represent different internet use features. The rows correspond to a combination of the time frame (30 days or 2 days) and device type (desktop or mobile) and the columns correspond to factors pertaining to individual differences  ( stress levels, gender, age, income). Red icons denote positive associations, while blue icons indicate negative associations. And the intensity of the color reflects the strength of the significance, with lighter icons denoting weak significance (0.05 $<$ p $<$ 0.1), and darker icons representing high significance (p $<$ 0.05).}
  \label{results_conceptual}
\end{figure}

Our results show that various internet usage behaviors are associated with stress, both positively and negatively. These associations differ across device type, time frame, and socio-demographic groups. Figure~\ref{results_conceptual} summarizes these patterns and their variation across panelist subgroups. 
In the following sections, we discuss the internet use features that have shown positive (social media, entertainment, shopping, games, and time of internet use), negative (adult content, productivity, and news), and mixed (messaging, and screen-time) associations with stress. 

\subsection{Positive Associations}
\subsubsection{Social Media}

People spend a substantial portion of their time online on social media platforms \citep{Statista2024}. In our dataset, social media accounted for approximately 23\% of total usage in mobile data and 15\% in desktop data. Therefore, understanding social media's relationship with stress is increasingly important as it continues to occupy a large share of individuals' internet activity. 

Prior research shows that social media can contribute to stress, act as a resource, or function as a coping tool \citep{Wolfers2022}. Factors such as fear of missing out (FOMO) \citep{Beyens2016}, appearance-related pressure \citep{berg2020}, and communication overload \citep{Chen2013} have been linked to increased distress. At the same time, other studies have highlighted its potential to buffer stress and offer social support in specific contexts \citep{Hampton2015, Rus2018, vanIngen2016}.

In our results, whenever significant, social media was consistently positively associated with stress across subgroups and device types. This association was significant for the high-stress subgroup in the 30-day mobile data, the Tier I income group in the 30-day desktop data, and the 30–45 age group in the 2-day desktop data (Fig~\ref{results_conceptual}). Although we cannot definitively determine whether social media contributes to or alleviates stress, these patterns could suggest that it may be used as a coping strategy for these groups.

Prior work has identified social media as a space for various coping mechanisms \citep{vanIngen2016}. Our findings in the middle-aged group potentially reflect this pattern, aligning with earlier research \citep{Wolfers2020}, possibly due to the support accessed through these platforms. This is consistent with studies linking social media use, particularly Facebook, to support-seeking behavior \citep{Brailovskaia2019}.

Overall, our results highlight the role of social media in shaping stress experiences. While our study, combining fine-grained web data with a longitudinal design and contextual framework, strengthens this interpretation, further research is needed to disentangle the causal directions of the observed associations and social media's role--stressor, resource, or coping tool.

\subsubsection{Entertainment}

According to a recent survey in Germany, people spend an average of 203 minutes per day watching content online \citep{Statista2024}. In our data, entertainment usage accounted for 9\% of total usage on mobile devices and 14\% on desktop devices.

Prior studies have reported a positive correlation between entertainment usage and stress \citep{khalili2019each, Shen2019, Aghababian2021, Alimoradi2022}. Further, activities such as watching content or listening to music have also been identified as common coping strategies for managing stress \citep{Boursier2021, Nabi2017, Hunter2009}. In our findings, entertainment usage was positively associated with stress in both desktop and mobile data. For mobile users, this association was consistent across both the 30-day and 2-day periods for the 30–45 age group. In the 2-day desktop data, a weak positive association was observed for the 18–30 age group.

These results suggest that entertainment usage (similar to social media) may serve as a coping mechanism, especially among younger users in our data, compared to older individuals. Prior research indicates that younger individuals are more likely to engage in binge-watching as a way to regulate emotions \citep{Boursier2021}. However, there is a lack of detailed research on this association \citep{Alimoradi2022}, and future studies should further explore the types of content consumed and their relationship to mental health.

\subsubsection{Online Shopping}

Online shopping has grown substantially in recent years, particularly with the rise of smartphones, which now account for 80\% of all retail visits \citep{Statista}. Previous studies have linked compulsive buying-shopping disorder (CBSD) to higher levels of stress \citep{Thomas2024, Li2022, Maraz2022, Tarka2022, Zheng2020, Singh2015}. At the same time, shopping has also been identified as a way to relieve stress \citep{Hama2001, Maharani2023}.

In our results, shopping was predominantly positively associated with stress across mobile and desktop data, in various time frames, and subgroups (see Figure~\ref{results_conceptual}), aligning with previous findings. A negative association was observed only for the 30–45 age group in the 2-day mobile data, suggesting that for this group, shopping may serve as a short-term stress reliever, or alternatively, that individuals under stress may avoid shopping. The latter behavior would align with prior research that has shown that, in middle-aged adults, stress can lead to increased saving behavior \citep{Durante2016}. 

This association was more pronounced in mobile data, suggesting that people may use mobile phones to cope with stress due to their easier accessibility. It was also consistent across both mobile and desktop data for the Tier IV income group, supporting earlier findings that higher-income individuals may use shopping as a way to cope with stress \citep{Maraz2022}.
In the 30-day mobile data, this positive association was observed among male participants but not among females. Prior studies show that males are more likely to experience negative emotions related to shopping \citep{Gallagher2017}, while some studies report no gender differences in online shopping addiction tendencies \citep{Li2022}.

These findings highlight how shopping has become embedded in daily life and its potential influence on stress levels. Future research should explore different types of shopping (hedonic vs. utilitarian) and their impact on mental health. Another direction could be to examine the time spent on shopping compared to actual purchases made after completing the payment process, and how these different shopping behaviors are associated with stress levels.


\subsubsection{Gaming}

Gaming has become a widespread daily habit, with the global market projected to reach US \$522.46 billion by 2025 \citep{Statista2025}. Previous research presents mixed findings: while gaming has been positively linked to stress and lower psychosocial well-being, with stress being a known precursor to pathological gaming \citep{MUN2023107767, PARK2024108002, Lemmens2011}, it has also been identified as a stress reliever and coping mechanism \citep{Whitbourne2013, Desai2021, Lee2023, pori2018}.

In our results, gaming usage was positively associated with stress in the 30-day mobile data for the high-stress subgroup. In the 2-day mobile data, positive associations were observed for the low-stress group, users aged 60+, and those in the Tier II income range. A negative association was found for the Tier V income group. These findings align with previous research suggesting that gaming may intensify stress in already stressed individuals \citep{Snodgrass2014}, while older adults often use gaming as a way to relieve stress \citep{Seer2023}. The contrasting patterns across income groups may reflect differences in usage intensity, as lower-income individuals tend to engage more frequently in gaming than those with higher incomes \citep{Engelsttter2022}.

We did not find any significant associations in the desktop data, which may be due to lower gaming activity on desktop browsers compared to mobile apps—gaming accounted for only 4.7\% of desktop usage versus 16.8\% in mobile data. Future research should examine the type of gaming, such as games that require active cognitive engagement (e.g., first-person shooter games) versus low cognitive requirement arcade games, and how these different types of games relate to stress.

\subsubsection{Timing of Online Activity}

Individuals tend to have preferences for the timing of both online and offline activities throughout the day \citep{aledavood2015digital, luong2023impact}.
Prior research has shown that increased nighttime smartphone use is linked to higher perceived stress \citep{Dissing2022}. Nighttime use has also been associated with reduced sleep duration \citep{Schrempft2024, Thomee2010} and poorer sleep quality \citep{Siebers2024, Luqman2021}, both of which are shown to impact mental health negatively \citep{Thome2011, Thome2012, Price2023}.

In our results, the daytime–nighttime difference feature was mainly positively associated with perceived stress across various groups. A negative association was observed only in the 60+ age group in the 30-day mobile data. Positive associations were more common in the shorter time frame, particularly in the 2-day mobile data, for groups including high-stress individuals, females, those aged 30–45, and the Tier IV income group (see Figure~\ref{results_conceptual}).

Our results suggest that more daytime use, compared to nighttime, is positively associated with stress, which contrasts with some earlier studies showing a stronger link between nighttime use and stress \citep{Dissing2022}. 
The internet use patterns can also be influenced by individuals' chronotypes, which in turn will affect preferred active hours \citep{Kortesoja2023}. 
Moreover, future studies could investigate panelists' bedtime and examine post-bedtime usage, as previous studies have shown that post-bedtime use, rather than nighttime use in general, has the most harmful effects on sleep quality \citep{Siebers2024}. The negative association observed in the 60+ age group may suggest that older individuals are more sensitive to late-night use and its link to stress. Further research is needed to better understand how daily temporal patterns of internet use relate to stress, particularly by considering users' chronotype and bedtime.


\subsection{Negative Associations}

\subsubsection{Adult Content}

A recent study estimates that around 90 million people may be affected by problematic adult content usage \citep{Bthe2024}. Research has shown that problematic adult content consumption can impact mental health, including associations with higher stress \citep{Laier2017, Altin2024}.

In our results, instead, adult content consumption was negatively associated with stress levels. This negative relationship was observed in the low-stress group (PSS score $<$ 14), and in the 18–30 and 60+ age groups within the 30-day desktop data. Some studies have reported no significant link between adult content use and psychological health \citep{Harper2016}, while others suggest it is often consumed as a form of leisure \citep{McCormack2017}. Our results might suggest that adult content serves as a stress buffer for our participants, as previous research has indicated stress and stress relief as common motivations for its use \citep{Bthe2021, Mubeen2022}. However, the positive associations observed in past studies were tied to problematic adult content consumption or addiction. This implies that while limited consumption may offer stress relief, excessive use could have detrimental effects. Future research should explore both the positive and negative impacts of adult content consumption on mental health, particularly in relation to the amount of consumption.

\subsubsection{Productivity}

Information and communication technology (ICT) has become a central part of daily work and study routines, particularly in the 21st century. Most of the work we do on ICT devices is connected to the internet, and their use for work and productivity has been linked to improved workplace efficiency \citep{Biagi2013, Buhari2024}. 
However, the effects of internet use on psychological health and stress are dual. While it has been associated with increased productivity, higher internet use for work-related tasks has also been linked to higher stress levels \citep{Afifi2018, Mark2012, Ninaus2015, Berg2018, 6720759}. Some interventional studies report no significant effects \citep{Berg2018}, while others suggest that using the internet for work and study can have a positive impact on mental health \citep{Hkby2016}.

In our results, increased use of productivity-related apps and domains was generally associated with lower perceived stress. In the 30-day mobile data, negative associations were observed among all participants, as well as within the male subgroup, the 30–45 age group, and the Tier III income group. In the 30-day desktop data, a similar negative association was found in the Tier I income group. A positive association appeared only once—in the 2-day desktop data for the Tier V income group.

Our findings suggest that stressed individuals may avoid productivity-related tasks by shifting their focus to other activities. This is supported by previous research showing a positive relationship between perceived stress and avoidant coping styles \citep{Allen2021}. Prior work has also noted that for younger individuals (up to 35 years) and older individuals (over 45 years), the relationship between ICT use and stress tends to be weak or nonexistent \citep{Berg2018}. For higher-income groups, the observed positive association may reflect greater work responsibilities that are more difficult to avoid, contributing to increased stress.

Overall, our findings highlight the need for further research into how productivity-related internet use influences stress. As prior studies have shown, avoidance behaviors can increase the risk of prolonged stress and other mental health challenges \citep{Holahan2005}. Future studies should explore how online interventions can effectively help users mitigate stress.

\subsubsection{News}

Informed citizens are a cornerstone of a well-functioning democracy and good governance. 
However, recent studies suggest that news, especially those with highly negative content, can adversely affect mental health and increase stress levels \citep{McLaughlin2023, Shaffer2024, Ladis2023, Kellerman2022}.

Our findings reveal a counterintuitive relationship between news engagement and stress: participants who spent more time consuming news tended to report lower stress levels. This association was more pronounced in the two days of data for both mobile and desktop users, and overall, it was stronger in desktop data across both time periods. One possible explanation is that individuals experiencing stress may avoid news altogether in the short term. Prior research supports this, showing that people under stress often disengage from news consumption \citep{Nguyen2021, Lindell2023, Mannell2022}. Other studies have found no significant link between news consumption and stress \citep{McNaughton, Lavelle2022}, while some suggest that positive and soft news content may improve mood \citep{Kelly2024} and well-being \citep{Boukes2017}.

Since our news category includes a range of sources, such as entertainment, sports, and politics, it is important to consider how different types of news relate to stress in the future. Future studies could examine the effects of specific news genres, as well as the influence of low-quality news sources and misinformation on mental health. 
Examining how consistent exposure to different news categories, sources, and misinformation influences stress responses over time could reveal insights into mental health risks and how media consumption habits shape psychological well-being.

\subsection{Mixed Associations}
\subsubsection{Messaging}

In 2021, an estimated 3.09 billion mobile phone users accessed the top messaging apps for communication, with this number projected to reach 3.51 billion by 2025 \citep{Laura}. The younger generation, in particular, remains connected with friends and family through messaging apps like WhatsApp and Telegram. Research has shown that messaging apps can be both positively associated with stress \citep{Coccia2016, Nimrod2020, Thome2007, Hurbean2022, Lin2021} and serve as stress reducers \citep{Melumad2021, Yau2021, Holtzman2017, Hooker2018}.

In our results, messaging usage was mostly positively associated with stress in mobile data. It was positively associated in the 30–45 age group and the 60+ age group, and negatively associated in the Tier I income group in the 30-day mobile data. In the 2-day mobile data, it was positively associated with females and those in the Tier III income group. In the 2-day desktop data, it was negatively associated with all participants and the 45–60 age group, while it was positively associated with the 60+ age group.

The key finding is that messaging use was predominantly positively associated with stress in mobile data and more negatively associated in desktop data. Previous research suggests that `checking behavior' (i.e., brief, repeated usage sessions) is more common in mobile use compared to desktop devices ~\citep{Oulasvirta2012}, and perhaps this type of repeated checking on mobile devices adds to the stress. 
We found that in the 60+ age group, messaging was positively associated with stress in both mobile and desktop data, which aligns with previous studies showing that older individuals experience higher stress from interpersonal communication \citep{Nimrod2020}. In the tier I income group for mobile data, messaging was negatively associated with stress. Studies show that people in lower-income groups tend to send more messages \citep{smith}. Previous research has also shown that messaging can be an effective tool for reducing depression, particularly among low-income individuals \citep{Aguilera2011}. 
Finally, in the 2-day mobile data, messaging was positively associated with stress in females but not males, reflecting previous research suggesting that frequent messaging is linked to mental health symptoms (externalizing and inattention) in females, but not in males \citep{George2020}.

Our results highlight how messaging can have a dual impact on stress, depending on the device, time period, and individual differences. Future work should explore how messaging with friends and family through apps, as well as interactions with strangers in chatrooms, can help alleviate stress and loneliness, ultimately improving well-being.

\subsubsection{Screen Time}

The internet plays a pervasive role in our daily lives, and the amount of time we spend online may have a detrimental impact on stress levels \citep{Yu2024,Thomee2010}. Research has shown that smartphone addiction is significantly linked to higher stress \citep{Elamin2024, Khan2023}, and the amount of time spent online is positively associated with stress \citep{Thomee2010} and other mental health issues \citep{Ding2024, Yue2020}.

In our results, total time spent online was positively associated with stress in the 30-day mobile data, aligning with previous studies \citep{Elamin2024, Khan2023, Thomee2010}. This association was significant for the high-stress group, individuals aged 60+, and those in income group IV. However, the relationship was mainly negative in both the 2-day and 30-day desktop data across different groups, with the exception of a positive association for the Tier IV income group in the 2-day desktop data, which mirrored the mobile data results.

Our findings suggest that increased time spent online on mobile devices may amplify stress in different contexts. For high-stress individuals, extended mobile phone usage is positively associated with stress, likely due to the constant accessibility of smartphones, making it harder to disconnect from them.  Prior studies have also found that smartphone users tend to experience higher levels of digital burnout compared to those using desktops or laptops \citep{Glda2022}. Additionally, older individuals in our study showed an increase in stress with more time spent online, consistent with previous research \citep{Ding2024}.

In contrast, desktop usage generally showed negative associations with stress, which may be due to the differences in accessibility of smartphones and desktop devices. Our data shows that time spent on desktop devices has higher variability, as reflected by the higher standard deviation(71.43 for desktop vs 66.86 for mobile) and coefficient of variation (0.97, CI=[0.93-1.02] for desktop vs 0.70, CI=[0.67-0.73] for mobile), compared to time spent on mobile devices.


Future research should further investigate device differences and the role mobile devices play in predicting stress levels.

\subsection{Implications}

Building on these findings, it is important to consider the implications for various stakeholders, including digital platform designers, health professionals, and end-users. 

For digital platform designers, these findings suggest opportunities to promote healthier use patterns. Features such as reminders to take breaks, tools to visualize usage habits, and reduced notifications during evening/nights could help users avoid stress-inducing behaviors. Additionally, adaptive designs that encourage daytime work engagement over late-night use could be particularly effective. Lastly, different design consideration would be needed for mobile or desktop devices to encourage healthier internet use.

For health professionals, understanding how digital behaviors relate to stress could offer new ways to support patients. For example, excessive gaming on mobile devices might indicate elevated stress. Health workers could include questions about internet habits in assessments and recommend tools that encourage healthier online behaviors. Notably, the low cost and high availability of web data could provide efficient tools to complement traditional monthly surveys for monitoring stress, helping to alleviate the burden on the already strained health care sector.


For individuals, these findings emphasize the importance of timing and purpose in digital habits. Being mindful of potentially stress-inducing activities, such as excessive shopping or gaming, can foster a healthier balance. Tools that track and suggest healthier patterns of internet use could assist users in managing their habits.

Overall, the relationship between stress and internet use is influenced by factors such as the type of activity, timing, and individual circumstances. This suggests that small, intentional changes in digital habits can help manage stress effectively. 


\subsection{Limitations}

The panelists we collected data from are gig workers who regularly participate in surveys, which may lead to behavioral differences compared to the general population. To address this issue, we employed rigorous preprocessing steps, including removing users who spent more than 25\% of their time on survey websites and excluding the bottom 20\% of users based on time spent.  Moreover, browsing data from these panels has been shown to prominently feature the most visited domains in Germany ~\citep{Kulshrestha2021}. Additionally, the privacy attitudes of web-tracked panelists have been found to be similar to those of non-web-tracked panelists from the same country ~\citep{Stier2020}. These findings support the reliability and suitability of the web tracking data for capturing individuals' internet usage behavior.
Also due to budget constraints, all panelists are from a single country, which may limit the generalizability of our findings due to cultural and behavioral differences.
Replicating this study in other countries or with a larger, more diverse sample could help improve generalizability. Additionally, 23\% of participants dropped out between the first and final survey waves. To mitigate this attrition, we initially took a larger sample and invited all panelists from the panel company to participate in our baseline survey. Furthermore, since our tracker only tracks browsing behavior through the browser, we lack complete data on desktop usage through desktop apps, which may limit our ability to fully capture differences between mobile and desktop use. Lastly, we used a validated self-report measure to assess stress, as inviting all participants to a lab setting was not feasible. Alternative approaches, such as physiological or real-time assessments, could provide more objective and detailed measures of stress.



\section{Conclusion}
Our study examines the relationship between internet use and perceived stress through a novel contextual framework that considers how, where, when, and by whom the internet is used. The findings indicate that internet use is associated with stress, and these associations differ across various usage contexts. Specifically, engagement with social media, online shopping, entertainment, and gaming is positively linked to higher stress levels. Notably, these activities have been identified in previous research as common coping mechanisms for stress, highlighting the need for future studies to examine whether such coping strategies alleviate stress or possibly exacerbate stress over time. However, productivity-related activities, news consumption, and adult content use are negatively associated with stress, suggesting they may either function as stress buffers or indicate avoidance behavior. Associations inferred from desktop data across different contextual dimensions are weaker than those inferred from mobile data, suggesting that the device type plays a role. 
In the short-term, news consumption is negatively associated for both mobile and desktop data. 
For individuals already experiencing high stress, increased time online on mobile phones, particularly on social media and gaming, is correlated with higher stress levels. Additionally, socio-demographic factors, especially income, have significant associations. These findings have important implications for the design of digital platforms, the development of mental health interventions, and the formation of healthier online habits. Future research should focus more specifically on particular web-based behaviors, such as news consumption and online shopping, and their effects on psychological well-being. It should also aim to establish causal links between internet use and stress, and further investigate the mechanisms underlying socio-demographic differences in these associations.

\section{CRediT authorship contribution statement}
Removed for double blind review.














\section{Ethics statement}

Our study was approved by our University's Research Ethics Committee (approval ID removed for double blind review). Data collection was conducted via a GDPR-compliant European company, and informed consent was obtained from participants for both the surveys and web traces datasets, with the option to withdraw consent at any time during or after the study. To protect participants' privacy, we implemented strict data privacy measures. The web dataset was anonymized by the panel company by removing personal information such as emails and usernames to prevent participant identification. Additionally, the dataset was stored and analyzed solely on the university’s server, with access restricted to the research team. We have made the anonymized data and code available to support the open-source community and spur further research at the intersection of internet use and well-being.

\section{Declaration of generative AI and AI-assisted technologies in the writing process}

During the preparation of this work the authors used ChatGPT for text proof reading (including spelling and grammar checks). After using this tool, the authors carefully reviewed and edited the content as necessary and take full responsibility for the final content of the published article.

\section{Funding}

Removed for double blind review.

\section{Declaration of competing interest}
The authors declare that they have no known competing financial interests or personal relationships that could influence the work reported in this paper.

\section{Acknowledgements}
Removed for double blind review.





\bibliographystyle{elsarticle-harv} 
\bibliography{ref}

\appendix

\section{Web data cleaning extended}

To ensure the integrity and reliability of the data, we applied a comprehensive preprocessing protocol. During this process, we observed that approximately 2–3\% of the visits in our dataset overlapped. Overlapping in this context means that a new visit was initiated before the previous visit from the same user had ended. While the exact cause of this overlap could not be determined, the most likely explanation is a potential issue with the tracking mechanism. To address this, we adjusted the duration of each visit: if a subsequent visit began before the prior visit ended, we recalculated the duration of the first visit as the time difference between its start time and the start of the overlapping visit.

\section{Mixed effect models results across contextual dimensions}
The following subsections present the results of both models across different groups and contexts. Model 1 includes coarse-grained features, while Model 2 incorporates fine-grained usage categories. For each predictor, we report estimates, confidence intervals (CI), and p-values, with statistically significant p-values bolded. Additionally, random effects, intraclass correlation coefficient (ICC), number of participants (N), and total observations are provided.

\subsection{Time period of data (\textit{When})}

\FloatBarrier

\begin{table}[H]
\centering
\begin{adjustbox}{max width=0.80\textwidth}
\begin{tabular}{l c r r c r r}
\toprule
\textbf{Predictors} & \multicolumn{3}{c}{Model 1} & \multicolumn{3}{c}{Model 2} \\
\cmidrule(lr){2-4} \cmidrule(lr){5-7}
& Estimate & CI & $p$ & Estimate & CI & $p$ \\
\midrule
Intercept & 20.67$^{***}$ & 18.89--22.44 & \textbf{$<$0.001} & 20.57$^{***}$ & 18.79--22.35 & \textbf{$<$0.001} \\
Survey wave & -0.11$^{*}$ & -0.21-- -0.01 & \textbf{0.024} & -0.11$^{*}$ & -0.20-- -0.01 & \textbf{0.030} \\
Gender & 2.02$^{***}$ & 0.97--3.06 & \textbf{$<$0.001} & 1.95$^{***}$ & 0.90--2.99 & \textbf{$<$0.001} \\
Age & -1.60$^{***}$ & -2.22-- -0.98 & \textbf{$<$0.001} & -1.56$^{***}$ & -2.18-- -0.93 & \textbf{$<$0.001} \\
Income & -1.17$^{***}$ & -1.58-- -0.76 & \textbf{$<$0.001} & -1.14$^{***}$ & -1.56-- -0.73 & \textbf{$<$0.001} \\
Total time spent online & 0.00 & -0.04--0.05 & 0.855 &  &  &  \\
Daytime nighttime difference & 0.05 & -0.03--0.12 & 0.203 & 0.05 & -0.03--0.12 & 0.223 \\
Time spent on entertainment &  &  &  & 0.01 & -0.12--0.15 & 0.840 \\
Time spent on social media &  &  &  & 0.02 & -0.11--0.15 & 0.811 \\
Time spent on messaging &  &  &  & 0.11 & -0.05--0.28 & 0.171 \\
Time spent on games &  &  &  & 0.01 & -0.08--0.10 & 0.780 \\
Time spent on shopping &  &  &  & -0.04 & -0.35--0.28 & 0.816 \\
Time spent on productivity &  &  &  & -0.05 & -0.29--0.18 & 0.648 \\
Time spent on news &  &  &  & -0.54$^{*}$ & -1.08-- -0.00 & \textbf{0.048} \\
\midrule
\multicolumn{7}{l}{\textbf{Random Effects}} \\
$\sigma^2$ & \multicolumn{3}{c}{12.02} & \multicolumn{3}{c}{12.02} \\
$\tau_{00}$ & \multicolumn{3}{c}{39.49$_{\text{pid}}$} & \multicolumn{3}{c}{39.43$_{\text{pid}}$} \\
ICC & \multicolumn{3}{c}{0.77} & \multicolumn{3}{c}{0.77} \\
N & \multicolumn{3}{c}{640$_{\text{pid}}$} & \multicolumn{3}{c}{640$_{\text{pid}}$} \\
Observations & \multicolumn{3}{c}{2141} & \multicolumn{3}{c}{2141} \\
\bottomrule
                \multicolumn{7}{l}{\textit{\textbf{* $p<0.05$ \quad ** $p<0.01$ \quad *** $p<0.001$}}} \\
                \end{tabular}
                \end{adjustbox}
                
\caption{Results from mixed-effects models for All Participants, based on 2 days of mobile usage data.}
\label{tab:mixed_effects_all_participants for last 2 of mobile data}
 \end{table}

\begin{table}[H]
\centering
\begin{adjustbox}{max width=0.80\textwidth}
\begin{tabular}{l c r r c r r}

\toprule
\textbf{Predictors} & \multicolumn{3}{c}{Model 1} & \multicolumn{3}{c}{Model 2} \\
\cmidrule(lr){2-4} \cmidrule(lr){5-7}
& Estimate & CI & $p$ & Estimate & CI & $p$ \\
\midrule
Intercept & 21.26$^{***}$ & 19.19--23.32 & \textbf{$<$0.001} & 21.23$^{***}$ & 19.17--23.29 & \textbf{$<$0.001} \\
Survey wave & -0.14$^{*}$ & -0.27-- -0.01 & \textbf{0.036} & -0.12 & -0.25--0.01 & 0.074 \\
Gender & 1.53$^{*}$ & 0.33--2.74 & \textbf{0.013} & 1.47$^{*}$ & 0.26--2.68 & \textbf{0.017} \\
Age & -1.96$^{***}$ & -2.68-- -1.24 & \textbf{$<$0.001} & -1.92$^{***}$ & -2.65-- -1.19 & \textbf{$<$0.001} \\
Income & -1.04$^{***}$ & -1.50-- -0.58 & \textbf{$<$0.001} & -1.02$^{***}$ & -1.48-- -0.56 & \textbf{$<$0.001} \\
Total time spent online & -0.01 & -0.08--0.05 & 0.695 &  &  &  \\
\end{tabular}
                \end{adjustbox}
                
\caption{Results from mixed-effects models for All Participants, based on 2 days of desktop usage data.}
\label{tab:mixed_effects_all_participants for last 2 of desktop data}
                \end{table}


\subsection{Baseline Stress Levels}


\begin{table}[H]
\centering
\begin{adjustbox}{max width=0.80\textwidth}
\begin{tabular}{l c r r c r r}
\toprule
\textbf{Predictors} & \multicolumn{3}{c}{Model 1} & \multicolumn{3}{c}{Model 2} \\
\cmidrule(lr){2-4} \cmidrule(lr){5-7}
& Estimate & CI & $p$ & Estimate & CI & $p$ \\
\midrule
Intercept & 29.70$^{***}$ & 27.97--31.43 & \textbf{$<$0.001} & 29.75$^{***}$ & 27.97--31.52 & \textbf{$<$0.001} \\
Survey wave & 0.09 & -0.08--0.27 & 0.305 & 0.10 & -0.09--0.28 & 0.298 \\
Gender & -0.42 & -1.32--0.48 & 0.357 & -0.30 & -1.23--0.64 & 0.533 \\
Age & -0.35 & -0.90--0.20 & 0.212 & -0.36 & -0.92--0.21 & 0.217 \\
Income & -0.50$^{**}$ & -0.88-- -0.12 & \textbf{0.010} & -0.52$^{*}$ & -0.91-- -0.12 & \textbf{0.010} \\
Total time spent online & 0.01$^{***}$ & 0.00--0.02 & \textbf{$<$0.001} &  &  &  \\
Daytime nighttime difference & 0.01 & -0.01--0.02 & 0.241 & 0.01 & -0.01--0.02 & 0.424 \\
Time spent on entertainment &  &  &  & 0.00 & -0.02--0.02 & 0.826 \\
Time spent on social media &  &  &  & 0.02$^{*}$ & 0.00--0.03 & \textbf{0.021} \\
Time spent on messaging &  &  &  & 0.00 & -0.03--0.03 & 0.846 \\
Time spent on games &  &  &  & 0.01$^{*}$ & 0.00--0.02 & \textbf{0.021} \\
Time spent on shopping &  &  &  & -0.01 & -0.07--0.05 & 0.697 \\
Time spent on productivity &  &  &  & 0.02 & -0.02--0.07 & 0.339 \\
Time spent on news &  &  &  & 0.09 & -0.02--0.20 & 0.102 \\
\midrule
\multicolumn{7}{l}{\textbf{Random Effects}} \\
$\sigma^2$ & \multicolumn{3}{c}{3.49} & \multicolumn{3}{c}{3.55} \\
$\tau_{00}$ & \multicolumn{3}{c}{2.33$_{\text{pid}}$} & \multicolumn{3}{c}{2.27$_{\text{pid}}$} \\
ICC & \multicolumn{3}{c}{0.40} & \multicolumn{3}{c}{0.39} \\
N & \multicolumn{3}{c}{94$_{\text{pid}}$} & \multicolumn{3}{c}{94$_{\text{pid}}$} \\
Observations & \multicolumn{3}{c}{197} & \multicolumn{3}{c}{197} \\
\bottomrule
                \multicolumn{7}{l}{\textit{\textbf{* $p<0.05$ \quad ** $p<0.01$ \quad *** $p<0.001$}}} \\
                \end{tabular}
                \end{adjustbox}
                
\caption{Results from mixed-effects models for High Stress, based on 30 days of mobile usage data.}
\label{tab:mixed_effects_high_stress for last 30 of mobile data}
                \end{table}
                
\begin{table}[H]
\centering
\begin{adjustbox}{max width=0.80\textwidth}
\begin{tabular}{l c r r c r r}

\toprule
\textbf{Predictors} & \multicolumn{3}{c}{Model 1} & \multicolumn{3}{c}{Model 2} \\
\cmidrule(lr){2-4} \cmidrule(lr){5-7}
& Estimate & CI & $p$ & Estimate & CI & $p$ \\
\midrule
Intercept & 32.53$^{***}$ & 30.04--35.01 & \textbf{$<$0.001} & 32.48$^{***}$ & 29.89--35.07 & \textbf{$<$0.001} \\
Survey wave & 0.04 & -0.21--0.30 & 0.756 & 0.05 & -0.21--0.32 & 0.703 \\
Gender & -0.72 & -2.11--0.67 & 0.309 & -0.99 & -2.49--0.52 & 0.199 \\
Age & -0.80 & -1.67--0.08 & 0.074 & -0.68 & -1.62--0.26 & 0.156 \\
Income & -0.43 & -0.99--0.12 & 0.127 & -0.37 & -0.95--0.21 & 0.211 \\
Total time spent online & -0.00 & -0.01--0.01 & 0.593 &  &  &  \\
Daytime nighttime difference & -0.01 & -0.03--0.01 & 0.453 & -0.00 & -0.02--0.02 & 0.946 \\
Time spent on entertainment &  &  &  & 0.02 & -0.01--0.05 & 0.205 \\
Time spent on adult content &  &  &  & -0.03 & -0.13--0.08 & 0.593 \\
Time spent on social media &  &  &  & 0.00 & -0.02--0.02 & 0.975 \\
Time spent on messaging &  &  &  & 0.01 & -0.27--0.29 & 0.931 \\
Time spent on games &  &  &  & 0.00 & -0.03--0.04 & 0.839 \\
Time spent on shopping &  &  &  & -0.01 & -0.06--0.05 & 0.785 \\
Time spent on productivity &  &  &  & -0.04 & -0.09--0.01 & 0.133 \\
Time spent on news &  &  &  & -0.00 & -0.11--0.10 & 0.959 \\
\midrule
\multicolumn{7}{l}{\textbf{Random Effects}} \\
$\sigma^2$ & \multicolumn{3}{c}{4.05} & \multicolumn{3}{c}{4.20} \\
$\tau_{00}$ & \multicolumn{3}{c}{4.82$_{\text{pid}}$} & \multicolumn{3}{c}{4.96$_{\text{pid}}$} \\
ICC & \multicolumn{3}{c}{0.54} & \multicolumn{3}{c}{0.54} \\
N & \multicolumn{3}{c}{63$_{\text{pid}}$} & \multicolumn{3}{c}{63$_{\text{pid}}$} \\
Observations & \multicolumn{3}{c}{132} & \multicolumn{3}{c}{132} \\
\bottomrule
                \multicolumn{7}{l}{\textit{\textbf{* $p<0.05$ \quad ** $p<0.01$ \quad *** $p<0.001$}}} \\
                \end{tabular}
                \end{adjustbox}
                
\caption{Results from mixed-effects models for High Stress, based on 30 days of desktop usage data.}
\label{tab:mixed_effects_high_stress for last 30 of desktop data}
                \end{table}

\begin{table}[H]
\centering
\begin{adjustbox}{max width=0.80\textwidth}
\begin{tabular}{l c r r c r r}
\toprule
\textbf{Predictors} & \multicolumn{3}{c}{Model 1} & \multicolumn{3}{c}{Model 2} \\
\cmidrule(lr){2-4} \cmidrule(lr){5-7}
& Estimate & CI & $p$ & Estimate & CI & $p$ \\
\midrule
Intercept & 31.83$^{***}$ & 30.02--33.63 & \textbf{$<$0.001} & 31.68$^{***}$ & 29.88--33.48 & \textbf{$<$0.001} \\
Survey wave & -0.03 & -0.22--0.17 & 0.799 & -0.02 & -0.22--0.18 & 0.868 \\
Gender & -0.73 & -1.75--0.30 & 0.166 & -0.66 & -1.70--0.38 & 0.215 \\
Age & -0.27 & -0.88--0.34 & 0.389 & -0.28 & -0.90--0.34 & 0.373 \\
Income & -0.67$^{**}$ & -1.11-- -0.24 & \textbf{0.002} & -0.64$^{**}$ & -1.07-- -0.21 & \textbf{0.004} \\
Total time spent online & 0.01 & -0.06--0.08 & 0.797 &  &  &  \\
Daytime nighttime difference & 0.11 & -0.01--0.23 & 0.075 & 0.09 & -0.04--0.22 & 0.154 \\
Time spent on entertainment &  &  &  & -0.04 & -0.24--0.17 & 0.730 \\
Time spent on social media &  &  &  & 0.04 & -0.15--0.23 & 0.686 \\
Time spent on messaging &  &  &  & -0.03 & -0.21--0.16 & 0.784 \\
Time spent on games &  &  &  & 0.06 & -0.09--0.20 & 0.447 \\
Time spent on shopping &  &  &  & 0.01 & -0.59--0.61 & 0.972 \\
Time spent on productivity &  &  &  & -0.14 & -0.63--0.35 & 0.577 \\
Time spent on news &  &  &  & 0.83 & -0.86--2.51 & 0.338 \\
\midrule
\multicolumn{7}{l}{\textbf{Random Effects}} \\
$\sigma^2$ & \multicolumn{3}{c}{2.92} & \multicolumn{3}{c}{3.11} \\
$\tau_{00}$ & \multicolumn{3}{c}{3.92$_{\text{pid}}$} & \multicolumn{3}{c}{3.68$_{\text{pid}}$} \\
ICC & \multicolumn{3}{c}{0.57} & \multicolumn{3}{c}{0.54} \\
N & \multicolumn{3}{c}{91$_{\text{pid}}$} & \multicolumn{3}{c}{91$_{\text{pid}}$} \\
Observations & \multicolumn{3}{c}{171} & \multicolumn{3}{c}{171} \\
\bottomrule
                \multicolumn{7}{l}{\textit{\textbf{* $p<0.05$ \quad ** $p<0.01$ \quad *** $p<0.001$}}} \\
                \end{tabular}
                \end{adjustbox}
                
\caption{Results from mixed-effects models for High Stress, based on 2 days of mobile usage data.}
\label{tab:mixed_effects_high_stress for last 2 of mobile data}
                \end{table}
                
\begin{table}[H]
\centering
\begin{adjustbox}{max width=0.80\textwidth}
\begin{tabular}{l c r r c r r}

\toprule
\textbf{Predictors} & \multicolumn{3}{c}{Model 1} & \multicolumn{3}{c}{Model 2} \\
\cmidrule(lr){2-4} \cmidrule(lr){5-7}
& Estimate & CI & $p$ & Estimate & CI & $p$ \\
\midrule
Intercept & 33.56$^{***}$ & 30.68--36.44 & \textbf{$<$0.001} & 33.81$^{***}$ & 30.76--36.85 & \textbf{$<$0.001} \\
Survey wave & -0.05 & -0.32--0.23 & 0.749 & -0.07 & -0.38--0.23 & 0.634 \\
Gender & -1.29 & -2.84--0.27 & 0.105 & -1.47 & -3.20--0.26 & 0.096 \\
Age & -1.04$^{*}$ & -2.04-- -0.03 & \textbf{0.043} & -1.02 & -2.06--0.03 & 0.057 \\
Income & -0.26 & -0.89--0.36 & 0.407 & -0.31 & -0.99--0.36 & 0.365 \\
Total time spent online & -0.06 & -0.17--0.05 & 0.278 &  &  &  \\
Daytime nighttime difference & 0.13 & -0.05--0.32 & 0.155 & 0.15 & -0.05--0.36 & 0.145 \\
Time spent on entertainment &  &  &  & -0.04 & -0.21--0.14 & 0.692 \\
Time spent on adult content &  &  &  & -0.73 & -1.81--0.35 & 0.183 \\
Time spent on social media &  &  &  & -0.02 & -0.33--0.29 & 0.898 \\
Time spent on messaging &  &  &  & 1.02 & -1.74--3.78 & 0.468 \\
Time spent on games &  &  &  & -0.00 & -0.43--0.43 & 0.996 \\
Time spent on shopping &  &  &  & -0.33 & -0.89--0.23 & 0.254 \\
Time spent on productivity &  &  &  & -0.12 & -0.65--0.40 & 0.648 \\
Time spent on news &  &  &  & 0.16 & -0.81--1.14 & 0.741 \\
\midrule
\multicolumn{7}{l}{\textbf{Random Effects}} \\
$\sigma^2$ & \multicolumn{3}{c}{4.49} & \multicolumn{3}{c}{4.71} \\
$\tau_{00}$ & \multicolumn{3}{c}{5.17$_{\text{pid}}$} & \multicolumn{3}{c}{5.33$_{\text{pid}}$} \\
ICC & \multicolumn{3}{c}{0.54} & \multicolumn{3}{c}{0.53} \\
N & \multicolumn{3}{c}{55$_{\text{pid}}$} & \multicolumn{3}{c}{55$_{\text{pid}}$} \\
Observations & \multicolumn{3}{c}{106} & \multicolumn{3}{c}{106} \\
\bottomrule
                \multicolumn{7}{l}{\textit{\textbf{* $p<0.05$ \quad ** $p<0.01$ \quad *** $p<0.001$}}} \\
                \end{tabular}
                \end{adjustbox}
                
\caption{Results from mixed-effects models for High Stress, based on 2 days of desktop usage data.}
\label{tab:mixed_effects_high_stress for last 2 of desktop data}
                \end{table}

\begin{table}[H]
\centering
\begin{adjustbox}{max width=0.80\textwidth}
\begin{tabular}{l c r r c r r}
\toprule
\textbf{Predictors} & \multicolumn{3}{c}{Model 1} & \multicolumn{3}{c}{Model 2} \\
\cmidrule(lr){2-4} \cmidrule(lr){5-7}
& Estimate & CI & $p$ & Estimate & CI & $p$ \\
\midrule
Intercept & 10.63$^{***}$ & 9.39--11.88 & \textbf{$<$0.001} & 10.39$^{***}$ & 9.13--11.65 & \textbf{$<$0.001} \\
Survey wave & -0.20$^{***}$ & -0.30-- -0.11 & \textbf{$<$0.001} & -0.20$^{***}$ & -0.29-- -0.10 & \textbf{$<$0.001} \\
Gender & 1.19$^{***}$ & 0.55--1.83 & \textbf{$<$0.001} & 1.08$^{**}$ & 0.42--1.74 & \textbf{0.001} \\
Age & -0.55$^{**}$ & -0.94-- -0.17 & \textbf{0.005} & -0.54$^{**}$ & -0.93-- -0.15 & \textbf{0.007} \\
Income & -0.12 & -0.37--0.13 & 0.345 & -0.11 & -0.36--0.15 & 0.414 \\
Total time spent online & -0.00 & -0.01--0.00 & 0.335 &  &  &  \\
Daytime nighttime difference & 0.00 & -0.01--0.01 & 0.737 & -0.00 & -0.01--0.01 & 0.923 \\
Time spent on entertainment &  &  &  & -0.01 & -0.03--0.00 & 0.124 \\
Time spent on social media &  &  &  & 0.01 & -0.01--0.02 & 0.223 \\
Time spent on messaging &  &  &  & 0.00 & -0.02--0.02 & 0.808 \\
Time spent on games &  &  &  & -0.00 & -0.01--0.01 & 0.995 \\
Time spent on shopping &  &  &  & 0.02 & -0.02--0.06 & 0.288 \\
Time spent on productivity &  &  &  & -0.02 & -0.05--0.01 & 0.141 \\
Time spent on news &  &  &  & 0.00 & -0.04--0.05 & 0.866 \\
\midrule
\multicolumn{7}{l}{\textbf{Random Effects}} \\
$\sigma^2$ & \multicolumn{3}{c}{5.66} & \multicolumn{3}{c}{5.65} \\
$\tau_{00}$ & \multicolumn{3}{c}{6.12$_{\text{pid}}$} & \multicolumn{3}{c}{6.14$_{\text{pid}}$} \\
ICC & \multicolumn{3}{c}{0.52} & \multicolumn{3}{c}{0.52} \\
N & \multicolumn{3}{c}{334$_{\text{pid}}$} & \multicolumn{3}{c}{334$_{\text{pid}}$} \\
Observations & \multicolumn{3}{c}{1011} & \multicolumn{3}{c}{1011} \\
\bottomrule
                \multicolumn{7}{l}{\textit{\textbf{* $p<0.05$ \quad ** $p<0.01$ \quad *** $p<0.001$}}} \\
                \end{tabular}
                \end{adjustbox}
                
\caption{Results from mixed-effects models for Low Stress, based on 30 days of mobile usage data.}
\label{tab:mixed_effects_low_stress for last 30 of mobile data}
                \end{table}
                
\begin{table}[htbp]
\centering
\begin{adjustbox}{max width=0.80\textwidth}
\begin{tabular}{l c r r c r r}

\toprule
\textbf{Predictors} & \multicolumn{3}{c}{Model 1} & \multicolumn{3}{c}{Model 2} \\
\cmidrule(lr){2-4} \cmidrule(lr){5-7}
& Estimate & CI & $p$ & Estimate & CI & $p$ \\
\midrule
Intercept & 10.27$^{***}$ & 8.80--11.75 & \textbf{$<$0.001} & 10.25$^{***}$ & 8.77--11.73 & \textbf{$<$0.001} \\
Survey wave & -0.14$^{**}$ & -0.25-- -0.04 & \textbf{0.009} & -0.15$^{**}$ & -0.26-- -0.04 & \textbf{0.008} \\
Gender & 0.37 & -0.41--1.15 & 0.355 & 0.36 & -0.43--1.15 & 0.376 \\
Age & -0.56$^{*}$ & -1.05-- -0.07 & \textbf{0.024} & -0.57$^{*}$ & -1.07-- -0.07 & \textbf{0.026} \\
Income & 0.02 & -0.28--0.33 & 0.882 & 0.01 & -0.30--0.33 & 0.935 \\
Total time spent online & -0.00 & -0.01--0.00 & 0.119 &  &  &  \\
Daytime nighttime difference & 0.00 & -0.00--0.01 & 0.263 & 0.01 & -0.00--0.01 & 0.232 \\
Time spent on entertainment &  &  &  & -0.00 & -0.01--0.00 & 0.341 \\
Time spent on adult content &  &  &  & -0.02 & -0.04--0.00 & 0.065 \\
Time spent on social media &  &  &  & -0.01 & -0.02--0.01 & 0.409 \\
Time spent on messaging &  &  &  & 0.02 & -0.01--0.05 & 0.276 \\
Time spent on games &  &  &  & -0.01 & -0.04--0.01 & 0.414 \\
Time spent on shopping &  &  &  & -0.00 & -0.04--0.03 & 0.936 \\
Time spent on productivity &  &  &  & 0.00 & -0.02--0.02 & 0.951 \\
Time spent on news &  &  &  & -0.00 & -0.03--0.03 & 0.861 \\
\midrule
\multicolumn{7}{l}{\textbf{Random Effects}} \\
$\sigma^2$ & \multicolumn{3}{c}{4.93} & \multicolumn{3}{c}{4.93} \\
$\tau_{00}$ & \multicolumn{3}{c}{7.31$_{\text{pid}}$} & \multicolumn{3}{c}{7.42$_{\text{pid}}$} \\
ICC & \multicolumn{3}{c}{0.60} & \multicolumn{3}{c}{0.60} \\
N & \multicolumn{3}{c}{268$_{\text{pid}}$} & \multicolumn{3}{c}{268$_{\text{pid}}$} \\
Observations & \multicolumn{3}{c}{712} & \multicolumn{3}{c}{712} \\
\bottomrule
                \multicolumn{7}{l}{\textit{\textbf{* $p<0.05$ \quad ** $p<0.01$ \quad *** $p<0.001$}}} \\
                \end{tabular}
                \end{adjustbox}
                
\caption{Results from mixed-effects models for Low Stress, based on 30 days of desktop usage data.}
\label{tab:mixed_effects_low_stress for last 30 of desktop data}
                \end{table}

\begin{table}[H]
\centering
\begin{adjustbox}{max width=0.80\textwidth}
\begin{tabular}{l c r r c r r}
\toprule
\textbf{Predictors} & \multicolumn{3}{c}{Model 1} & \multicolumn{3}{c}{Model 2} \\
\cmidrule(lr){2-4} \cmidrule(lr){5-7}
& Estimate & CI & $p$ & Estimate & CI & $p$ \\
\midrule
Intercept & 10.25$^{***}$ & 8.95--11.55 & \textbf{$<$0.001} & 10.30$^{***}$ & 8.98--11.62 & \textbf{$<$0.001} \\
Survey wave & -0.24$^{***}$ & -0.34-- -0.13 & \textbf{$<$0.001} & -0.23$^{***}$ & -0.34-- -0.12 & \textbf{$<$0.001} \\
Gender & 1.18$^{***}$ & 0.51--1.84 & \textbf{$<$0.001} & 1.10$^{**}$ & 0.43--1.78 & \textbf{0.001} \\
Age & -0.52$^{*}$ & -0.93-- -0.12 & \textbf{0.011} & -0.55$^{**}$ & -0.97-- -0.14 & \textbf{0.008} \\
Income & -0.12 & -0.38--0.15 & 0.379 & -0.10 & -0.37--0.17 & 0.460 \\
Total time spent online & 0.01 & -0.05--0.06 & 0.813 &  &  &  \\
Daytime nighttime difference & 0.00 & -0.08--0.09 & 0.971 & -0.02 & -0.11--0.06 & 0.606 \\
Time spent on entertainment &  &  &  & -0.13 & -0.29--0.04 & 0.136 \\
Time spent on social media &  &  &  & -0.03 & -0.20--0.13 & 0.684 \\
Time spent on messaging &  &  &  & 0.09 & -0.14--0.32 & 0.453 \\
Time spent on games &  &  &  & 0.08 & -0.01--0.17 & 0.098 \\
Time spent on shopping &  &  &  & 0.04 & -0.28--0.37 & 0.799 \\
Time spent on productivity &  &  &  & 0.04 & -0.15--0.23 & 0.696 \\
Time spent on news &  &  &  & -0.30 & -0.81--0.22 & 0.257 \\
\midrule
\multicolumn{7}{l}{\textbf{Random Effects}} \\
$\sigma^2$ & \multicolumn{3}{c}{5.96} & \multicolumn{3}{c}{5.92} \\
$\tau_{00}$ & \multicolumn{3}{c}{6.03$_{\text{pid}}$} & \multicolumn{3}{c}{6.13$_{\text{pid}}$} \\
ICC & \multicolumn{3}{c}{0.50} & \multicolumn{3}{c}{0.51} \\
N & \multicolumn{3}{c}{323$_{\text{pid}}$} & \multicolumn{3}{c}{323$_{\text{pid}}$} \\
Observations & \multicolumn{3}{c}{849} & \multicolumn{3}{c}{849} \\
\bottomrule
                \multicolumn{7}{l}{\textit{\textbf{* $p<0.05$ \quad ** $p<0.01$ \quad *** $p<0.001$}}} \\
                \end{tabular}
                \end{adjustbox}
                
\caption{Results from mixed-effects models for Low Stress, based on 2 days of mobile usage data. }
\label{tab:mixed_effects_low_stress for last 2 of mobile data}
                \end{table}
                
\begin{table}[H]
\centering
\begin{adjustbox}{max width=0.80\textwidth}
\begin{tabular}{l c r r c r r}

\toprule
\textbf{Predictors} & \multicolumn{3}{c}{Model 1} & \multicolumn{3}{c}{Model 2} \\
\cmidrule(lr){2-4} \cmidrule(lr){5-7}
& Estimate & CI & $p$ & Estimate & CI & $p$ \\
\midrule
Intercept & 10.04$^{***}$ & 8.32--11.77 & \textbf{$<$0.001} & 10.01$^{***}$ & 8.28--11.75 & \textbf{$<$0.001} \\
Survey wave & -0.20$^{**}$ & -0.33-- -0.07 & \textbf{0.003} & -0.21$^{**}$ & -0.35-- -0.07 & \textbf{0.003} \\
Gender & 0.35 & -0.48--1.18 & 0.411 & 0.35 & -0.49--1.19 & 0.413 \\
Age & -0.10 & -0.66--0.46 & 0.720 & -0.10 & -0.67--0.47 & 0.729 \\
Income & -0.21 & -0.54--0.12 & 0.211 & -0.22 & -0.56--0.11 & 0.193 \\
Total time spent online & -0.07$^{*}$ & -0.14-- -0.00 & \textbf{0.038} &  &  &  \\
Daytime nighttime difference & 0.04 & -0.05--0.12 & 0.367 & 0.04 & -0.05--0.13 & 0.438 \\
Time spent on entertainment &  &  &  & -0.04 & -0.16--0.08 & 0.516 \\
Time spent on adult content &  &  &  & -0.11 & -0.31--0.10 & 0.310 \\
Time spent on social media &  &  &  & -0.10 & -0.31--0.10 & 0.316 \\
Time spent on messaging &  &  &  & -0.21 & -0.56--0.15 & 0.253 \\
Time spent on games &  &  &  & -0.24 & -0.57--0.09 & 0.155 \\
Time spent on shopping &  &  &  & -0.13 & -0.44--0.18 & 0.408 \\
Time spent on productivity &  &  &  & 0.07 & -0.14--0.28 & 0.504 \\
Time spent on news &  &  &  & -0.16 & -0.52--0.20 & 0.387 \\
\midrule
\multicolumn{7}{l}{\textbf{Random Effects}} \\
$\sigma^2$ & \multicolumn{3}{c}{4.86} & \multicolumn{3}{c}{4.88} \\
$\tau_{00}$ & \multicolumn{3}{c}{7.31$_{\text{pid}}$} & \multicolumn{3}{c}{7.43$_{\text{pid}}$} \\
ICC & \multicolumn{3}{c}{0.60} & \multicolumn{3}{c}{0.60} \\
N & \multicolumn{3}{c}{241$_{\text{pid}}$} & \multicolumn{3}{c}{241$_{\text{pid}}$} \\
Observations & \multicolumn{3}{c}{500} & \multicolumn{3}{c}{500} \\
\bottomrule
                \multicolumn{7}{l}{\textit{\textbf{* $p<0.05$ \quad ** $p<0.01$ \quad *** $p<0.001$}}} \\
                \end{tabular}
                \end{adjustbox}
                
\caption{Results from mixed-effects models for Low Stress, based on 2 days of desktop usage data.}
\label{tab:mixed_effects_low_stress for last 2 of desktop data}
                \end{table}

\subsection{Gender Differences}


\begin{table}[H]
\centering
\begin{adjustbox}{max width=0.80\textwidth}
\begin{tabular}{l c r r c r r}
\toprule
\textbf{Predictors} & \multicolumn{3}{c}{Model 1} & \multicolumn{3}{c}{Model 2} \\
\cmidrule(lr){2-4} \cmidrule(lr){5-7}
& Estimate & CI & $p$ & Estimate & CI & $p$ \\
\midrule
Intercept & 21.43$^{***}$ & 19.22--23.65 & \textbf{$<$0.001} & 21.14$^{***}$ & 18.92--23.35 & \textbf{$<$0.001} \\
Survey wave & -0.12$^{*}$ & -0.24-- -0.01 & \textbf{0.035} & -0.13$^{*}$ & -0.25-- -0.01 & \textbf{0.028} \\
Age & -1.81$^{***}$ & -2.65-- -0.97 & \textbf{$<$0.001} & -1.74$^{***}$ & -2.59-- -0.89 & \textbf{$<$0.001} \\
Income & -1.19$^{***}$ & -1.74-- -0.64 & \textbf{$<$0.001} & -1.12$^{***}$ & -1.67-- -0.57 & \textbf{$<$0.001} \\
Total time spent online & 0.00 & -0.01--0.01 & 0.803 &  &  &  \\
Daytime nighttime difference & -0.01 & -0.02--0.01 & 0.365 & -0.01 & -0.02--0.01 & 0.462 \\
Time spent on entertainment &  &  &  & -0.00 & -0.02--0.02 & 0.901 \\
Time spent on social media &  &  &  & 0.01 & -0.01--0.03 & 0.416 \\
Time spent on messaging &  &  &  & 0.01 & -0.02--0.03 & 0.615 \\
Time spent on games &  &  &  & 0.00 & -0.01--0.02 & 0.602 \\
Time spent on shopping &  &  &  & 0.07$^{*}$ & 0.01--0.14 & \textbf{0.022} \\
Time spent on productivity &  &  &  & -0.05$^{*}$ & -0.09-- -0.01 & \textbf{0.021} \\
Time spent on news &  &  &  & -0.04 & -0.11--0.03 & 0.241 \\
\midrule
\multicolumn{7}{l}{\textbf{Random Effects}} \\
$\sigma^2$ & \multicolumn{3}{c}{11.93} & \multicolumn{3}{c}{11.89} \\
$\tau_{00}$ & \multicolumn{3}{c}{37.51$_{\text{pid}}$} & \multicolumn{3}{c}{37.17$_{\text{pid}}$} \\
ICC & \multicolumn{3}{c}{0.76} & \multicolumn{3}{c}{0.76} \\
N & \multicolumn{3}{c}{334$_{\text{pid}}$} & \multicolumn{3}{c}{334$_{\text{pid}}$} \\
Observations & \multicolumn{3}{c}{1334} & \multicolumn{3}{c}{1334} \\
\bottomrule
                \multicolumn{7}{l}{\textit{\textbf{* $p<0.05$ \quad ** $p<0.01$ \quad *** $p<0.001$}}} \\
                \end{tabular}
                \end{adjustbox}
                
\caption{Results from mixed-effects models for Male, based on 30 days of mobile usage data.}
\label{tab:mixed_effects_male for last 30 of mobile data}
                \end{table}
                
\begin{table}[H]
\centering
\begin{adjustbox}{max width=0.80\textwidth}
\begin{tabular}{l c r r c r r}
\toprule
\textbf{Predictors} & \multicolumn{3}{c}{Model 1} & \multicolumn{3}{c}{Model 2} \\
\cmidrule(lr){2-4} \cmidrule(lr){5-7}
& Estimate & CI & $p$ & Estimate & CI & $p$ \\
\midrule
Intercept & 21.80$^{***}$ & 19.58--24.02 & \textbf{$<$0.001} & 21.94$^{***}$ & 19.68--24.19 & \textbf{$<$0.001} \\
Survey wave & -0.08 & -0.20--0.04 & 0.192 & -0.08 & -0.20--0.04 & 0.171 \\
Age & -1.41$^{***}$ & -2.21-- -0.62 & \textbf{$<$0.001} & -1.44$^{***}$ & -2.25-- -0.62 & \textbf{$<$0.001} \\
Income & -1.00$^{***}$ & -1.54-- -0.46 & \textbf{$<$0.001} & -1.01$^{***}$ & -1.56-- -0.45 & \textbf{$<$0.001} \\
Total time spent online & 0.00 & -0.00--0.01 & 0.324 &  &  &  \\
Daytime nighttime difference & -0.01 & -0.02--0.00 & 0.214 & -0.01 & -0.02--0.01 & 0.279 \\
Time spent on entertainment &  &  &  & 0.02 & -0.01--0.04 & 0.140 \\
Time spent on social media &  &  &  & -0.00 & -0.02--0.01 & 0.711 \\
Time spent on messaging &  &  &  & -0.00 & -0.03--0.02 & 0.889 \\
Time spent on games &  &  &  & 0.00 & -0.01--0.02 & 0.389 \\
Time spent on shopping &  &  &  & 0.02 & -0.03--0.07 & 0.466 \\
Time spent on productivity &  &  &  & -0.01 & -0.05--0.03 & 0.610 \\
Time spent on news &  &  &  & 0.03 & -0.10--0.16 & 0.676 \\
\midrule
\multicolumn{7}{l}{\textbf{Random Effects}} \\
$\sigma^2$ & \multicolumn{3}{c}{11.41} & \multicolumn{3}{c}{11.41} \\
$\tau_{00}$ & \multicolumn{3}{c}{35.76$_{\text{pid}}$} & \multicolumn{3}{c}{36.19$_{\text{pid}}$} \\
ICC & \multicolumn{3}{c}{0.76} & \multicolumn{3}{c}{0.76} \\
N & \multicolumn{3}{c}{322$_{\text{pid}}$} & \multicolumn{3}{c}{322$_{\text{pid}}$} \\
Observations & \multicolumn{3}{c}{1266} & \multicolumn{3}{c}{1266} \\
\bottomrule
                \multicolumn{7}{l}{\textit{\textbf{* $p<0.05$ \quad ** $p<0.01$ \quad *** $p<0.001$}}} \\
                \end{tabular}
                \end{adjustbox}
                
\caption{Results from mixed-effects models for Female, based on 30 days of mobile usage data.}
\label{tab:mixed_effects_female for last 30 of mobile data}
                \end{table}


\begin{table}[H]
\centering
\begin{adjustbox}{max width=0.80\textwidth}
\begin{tabular}{l c r r c r r}

\toprule
\textbf{Predictors} & \multicolumn{3}{c}{Model 1} & \multicolumn{3}{c}{Model 2} \\
\cmidrule(lr){2-4} \cmidrule(lr){5-7}
& Estimate & CI & $p$ & Estimate & CI & $p$ \\
\midrule
Intercept & 20.38$^{***}$ & 17.91--22.85 & \textbf{$<$0.001} & 20.16$^{***}$ & 17.70--22.62 & \textbf{$<$0.001} \\
Survey wave & -0.15$^{*}$ & -0.29-- -0.02 & \textbf{0.029} & -0.17$^{*}$ & -0.31-- -0.02 & \textbf{0.021} \\
Age & -1.52$^{**}$ & -2.47-- -0.57 & \textbf{0.002} & -1.50$^{**}$ & -2.47-- -0.54 & \textbf{0.002} \\
Income & -0.93$^{**}$ & -1.50-- -0.35 & \textbf{0.002} & -0.92$^{**}$ & -1.51-- -0.34 & \textbf{0.002} \\
Total time spent online & -0.00 & -0.01--0.00 & 0.359 &  &  &  \\
Daytime nighttime difference & -0.00 & -0.01--0.01 & 0.797 & -0.00 & -0.01--0.01 & 0.910 \\
Time spent on entertainment &  &  &  & -0.00 & -0.01--0.01 & 0.973 \\
Time spent on adult content &  &  &  & -0.01 & -0.03--0.01 & 0.230 \\
Time spent on social media &  &  &  & 0.01 & -0.02--0.04 & 0.677 \\
Time spent on messaging &  &  &  & 0.03 & -0.06--0.12 & 0.524 \\
Time spent on games &  &  &  & -0.03 & -0.07--0.02 & 0.276 \\
Time spent on shopping &  &  &  & 0.02 & -0.02--0.07 & 0.334 \\
Time spent on productivity &  &  &  & -0.00 & -0.03--0.03 & 0.762 \\
Time spent on news &  &  &  & -0.01 & -0.06--0.03 & 0.595 \\
\midrule
\multicolumn{7}{l}{\textbf{Random Effects}} \\
$\sigma^2$ & \multicolumn{3}{c}{11.23} & \multicolumn{3}{c}{11.27} \\
$\tau_{00}$ & \multicolumn{3}{c}{39.79$_{\text{pid}}$} & \multicolumn{3}{c}{39.96$_{\text{pid}}$} \\
ICC & \multicolumn{3}{c}{0.78} & \multicolumn{3}{c}{0.78} \\
N & \multicolumn{3}{c}{289$_{\text{pid}}$} & \multicolumn{3}{c}{289$_{\text{pid}}$} \\
Observations & \multicolumn{3}{c}{974} & \multicolumn{3}{c}{974} \\
\bottomrule
                \multicolumn{7}{l}{\textit{\textbf{* $p<0.05$ \quad ** $p<0.01$ \quad *** $p<0.001$}}} \\
                \end{tabular}
                \end{adjustbox}
                
\caption{Results from mixed-effects models for Male, based on 30 days of desktop usage data.}
\label{tab:mixed_effects_male for last 30 of desktop data}
                \end{table}
                
\begin{table}[H]
\centering
\begin{adjustbox}{max width=0.80\textwidth}
\begin{tabular}{l c r r c r r}

\toprule
\textbf{Predictors} & \multicolumn{3}{c}{Model 1} & \multicolumn{3}{c}{Model 2} \\
\cmidrule(lr){2-4} \cmidrule(lr){5-7}
& Estimate & CI & $p$ & Estimate & CI & $p$ \\
\midrule
Intercept & 22.76$^{***}$ & 20.10--25.43 & \textbf{$<$0.001} & 22.68$^{***}$ & 20.01--25.36 & \textbf{$<$0.001} \\
Survey wave & -0.11 & -0.28--0.05 & 0.162 & -0.12 & -0.28--0.04 & 0.146 \\
Age & -1.97$^{***}$ & -2.96-- -0.98 & \textbf{$<$0.001} & -1.99$^{***}$ & -3.00-- -0.98 & \textbf{$<$0.001} \\
Income & -0.98$^{**}$ & -1.65-- -0.31 & \textbf{0.004} & -0.98$^{**}$ & -1.66-- -0.30 & \textbf{0.005} \\
Total time spent online & -0.00 & -0.01--0.01 & 0.692 &  &  &  \\
Daytime nighttime difference & 0.01 & -0.00--0.03 & 0.152 & 0.01 & -0.00--0.03 & 0.105 \\
Time spent on entertainment &  &  &  & -0.01 & -0.04--0.01 & 0.226 \\
Time spent on adult content &  &  &  & -0.02 & -0.05--0.01 & 0.226 \\
Time spent on social media &  &  &  & -0.00 & -0.03--0.02 & 0.699 \\
Time spent on messaging &  &  &  & 0.00 & -0.06--0.06 & 0.973 \\
Time spent on games &  &  &  & 0.03 & -0.02--0.07 & 0.233 \\
Time spent on shopping &  &  &  & 0.04 & -0.01--0.09 & 0.147 \\
Time spent on productivity &  &  &  & -0.00 & -0.05--0.05 & 0.978 \\
Time spent on news &  &  &  & -0.05 & -0.12--0.02 & 0.146 \\
\midrule
\multicolumn{7}{l}{\textbf{Random Effects}} \\
$\sigma^2$ & \multicolumn{3}{c}{10.95} & \multicolumn{3}{c}{10.89} \\
$\tau_{00}$ & \multicolumn{3}{c}{41.82$_{\text{pid}}$} & \multicolumn{3}{c}{42.19$_{\text{pid}}$} \\
ICC & \multicolumn{3}{c}{0.79} & \multicolumn{3}{c}{0.79} \\
N & \multicolumn{3}{c}{237$_{\text{pid}}$} & \multicolumn{3}{c}{237$_{\text{pid}}$} \\
Observations & \multicolumn{3}{c}{739} & \multicolumn{3}{c}{739} \\
\bottomrule
                \multicolumn{7}{l}{\textit{\textbf{* $p<0.05$ \quad ** $p<0.01$ \quad *** $p<0.001$}}} \\
                \end{tabular}
                \end{adjustbox}
                
\caption{Results from mixed-effects models for Female, based on 30 days of desktop usage data.}
\label{tab:mixed_effects_female for last 30 of desktop data}
                \end{table}


\begin{table}[H]
\centering
\begin{adjustbox}{max width=0.80\textwidth}
\begin{tabular}{l c r r c r r}
\toprule
\textbf{Predictors} & \multicolumn{3}{c}{Model 1} & \multicolumn{3}{c}{Model 2} \\
\cmidrule(lr){2-4} \cmidrule(lr){5-7}
& Estimate & CI & $p$ & Estimate & CI & $p$ \\
\midrule
Intercept & 21.94$^{***}$ & 19.55--24.32 & \textbf{$<$0.001} & 21.88$^{***}$ & 19.50--24.27 & \textbf{$<$0.001} \\
Survey wave & -0.11 & -0.24--0.03 & 0.120 & -0.11 & -0.25--0.03 & 0.112 \\
Age & -2.12$^{***}$ & -3.01-- -1.23 & \textbf{$<$0.001} & -2.08$^{***}$ & -2.97-- -1.18 & \textbf{$<$0.001} \\
Income & -1.26$^{***}$ & -1.85-- -0.68 & \textbf{$<$0.001} & -1.23$^{***}$ & -1.81-- -0.64 & \textbf{$<$0.001} \\
Total time spent online & -0.02 & -0.09--0.05 & 0.576 &  &  &  \\
Daytime nighttime difference & 0.01 & -0.09--0.11 & 0.850 & 0.01 & -0.09--0.12 & 0.847 \\
Time spent on entertainment &  &  &  & -0.08 & -0.27--0.11 & 0.414 \\
Time spent on social media &  &  &  & 0.07 & -0.12--0.27 & 0.471 \\
Time spent on messaging &  &  &  & 0.03 & -0.19--0.25 & 0.789 \\
Time spent on games &  &  &  & 0.01 & -0.13--0.15 & 0.891 \\
Time spent on shopping &  &  &  & -0.23 & -0.74--0.27 & 0.365 \\
Time spent on productivity &  &  &  & -0.07 & -0.35--0.22 & 0.637 \\
Time spent on news &  &  &  & -0.58 & -1.23--0.07 & 0.079 \\
\midrule
\multicolumn{7}{l}{\textbf{Random Effects}} \\
$\sigma^2$ & \multicolumn{3}{c}{12.59} & \multicolumn{3}{c}{12.61} \\
$\tau_{00}$ & \multicolumn{3}{c}{41.22$_{\text{pid}}$} & \multicolumn{3}{c}{41.08$_{\text{pid}}$} \\
ICC & \multicolumn{3}{c}{0.77} & \multicolumn{3}{c}{0.77} \\
N & \multicolumn{3}{c}{335$_{\text{pid}}$} & \multicolumn{3}{c}{335$_{\text{pid}}$} \\
Observations & \multicolumn{3}{c}{1092} & \multicolumn{3}{c}{1092} \\
\bottomrule
                \multicolumn{7}{l}{\textit{\textbf{* $p<0.05$ \quad ** $p<0.01$ \quad *** $p<0.001$}}} \\
                \end{tabular}
                \end{adjustbox}
                
\caption{Results from mixed-effects models for Male, based on 2 days of mobile usage data.}
\label{tab:mixed_effects_male for last 2 of mobile data}
                \end{table}
                
\begin{table}[H]
\centering
\begin{adjustbox}{max width=0.80\textwidth}
\begin{tabular}{l c r r c r r}
\toprule
\textbf{Predictors} & \multicolumn{3}{c}{Model 1} & \multicolumn{3}{c}{Model 2} \\
\cmidrule(lr){2-4} \cmidrule(lr){5-7}
& Estimate & CI & $p$ & Estimate & CI & $p$ \\
\midrule
Intercept & 21.31$^{***}$ & 19.02--23.59 & \textbf{$<$0.001} & 21.19$^{***}$ & 18.89--23.49 & \textbf{$<$0.001} \\
Survey wave & -0.11 & -0.24--0.03 & 0.124 & -0.10 & -0.24--0.03 & 0.136 \\
Age & -1.08$^{*}$ & -1.93-- -0.22 & \textbf{0.014} & -1.07$^{*}$ & -1.93-- -0.20 & \textbf{0.016} \\
Income & -1.03$^{***}$ & -1.60-- -0.45 & \textbf{$<$0.001} & -1.02$^{***}$ & -1.60-- -0.44 & \textbf{$<$0.001} \\
Total time spent online & 0.03 & -0.04--0.09 & 0.422 &  &  &  \\
Daytime nighttime difference & 0.09 & -0.02--0.19 & 0.095 & 0.10 & -0.00--0.21 & 0.060 \\
Time spent on entertainment &  &  &  & 0.14 & -0.06--0.34 & 0.167 \\
Time spent on social media &  &  &  & -0.05 & -0.23--0.13 & 0.567 \\
Time spent on messaging &  &  &  & 0.23 & -0.02--0.48 & 0.075 \\
Time spent on games &  &  &  & 0.00 & -0.11--0.12 & 0.944 \\
Time spent on shopping &  &  &  & 0.09 & -0.32--0.50 & 0.668 \\
Time spent on productivity &  &  &  & -0.00 & -0.41--0.41 & 0.996 \\
Time spent on news &  &  &  & -0.37 & -1.37--0.62 & 0.462 \\
\midrule
\multicolumn{7}{l}{\textbf{Random Effects}} \\
$\sigma^2$ & \multicolumn{3}{c}{11.43} & \multicolumn{3}{c}{11.43} \\
$\tau_{00}$ & \multicolumn{3}{c}{37.55$_{\text{pid}}$} & \multicolumn{3}{c}{37.60$_{\text{pid}}$} \\
ICC & \multicolumn{3}{c}{0.77} & \multicolumn{3}{c}{0.77} \\
N & \multicolumn{3}{c}{305$_{\text{pid}}$} & \multicolumn{3}{c}{305$_{\text{pid}}$} \\
Observations & \multicolumn{3}{c}{1049} & \multicolumn{3}{c}{1049} \\
\bottomrule
                \multicolumn{7}{l}{\textit{\textbf{* $p<0.05$ \quad ** $p<0.01$ \quad *** $p<0.001$}}} \\
                \end{tabular}
                \end{adjustbox}
                
\caption{Results from mixed-effects models for Female, based on 2 days of mobile usage data.}
\label{tab:mixed_effects_female for last 2 of mobile data}
                \end{table}


\begin{table}[H]
\centering
\begin{adjustbox}{max width=0.80\textwidth}
\begin{tabular}{l c r r c r r}

\toprule
\textbf{Predictors} & \multicolumn{3}{c}{Model 1} & \multicolumn{3}{c}{Model 2} \\
\cmidrule(lr){2-4} \cmidrule(lr){5-7}
& Estimate & CI & $p$ & Estimate & CI & $p$ \\
\midrule
Intercept & 20.58$^{***}$ & 17.90--23.27 & \textbf{$<$0.001} & 20.42$^{***}$ & 17.73--23.11 & \textbf{$<$0.001} \\
Survey wave & -0.11 & -0.28--0.06 & 0.202 & -0.10 & -0.27--0.07 & 0.246 \\
Age & -1.67$^{**}$ & -2.68-- -0.66 & \textbf{0.001} & -1.62$^{**}$ & -2.63-- -0.60 & \textbf{0.002} \\
Income & -1.05$^{***}$ & -1.66-- -0.45 & \textbf{$<$0.001} & -1.02$^{**}$ & -1.64-- -0.41 & \textbf{0.001} \\
Total time spent online & -0.00 & -0.09--0.08 & 0.916 &  &  &  \\
Daytime nighttime difference & 0.07 & -0.04--0.19 & 0.195 & 0.07 & -0.04--0.19 & 0.230 \\
Time spent on entertainment &  &  &  & 0.06 & -0.08--0.19 & 0.412 \\
Time spent on adult content &  &  &  & -0.09 & -0.33--0.16 & 0.484 \\
Time spent on social media &  &  &  & 0.13 & -0.22--0.48 & 0.458 \\
Time spent on messaging &  &  &  & -0.63 & -1.64--0.38 & 0.224 \\
Time spent on games &  &  &  & -0.32 & -0.73--0.09 & 0.128 \\
Time spent on shopping &  &  &  & -0.16 & -0.58--0.26 & 0.461 \\
Time spent on productivity &  &  &  & 0.21 & -0.12--0.53 & 0.211 \\
Time spent on news &  &  &  & -0.52$^{*}$ & -1.02-- -0.01 & \textbf{0.044} \\
\midrule
\multicolumn{7}{l}{\textbf{Random Effects}} \\
$\sigma^2$ & \multicolumn{3}{c}{10.43} & \multicolumn{3}{c}{10.29} \\
$\tau_{00}$ & \multicolumn{3}{c}{39.39$_{\text{pid}}$} & \multicolumn{3}{c}{39.67$_{\text{pid}}$} \\
ICC & \multicolumn{3}{c}{0.79} & \multicolumn{3}{c}{0.79} \\
N & \multicolumn{3}{c}{270$_{\text{pid}}$} & \multicolumn{3}{c}{270$_{\text{pid}}$} \\
Observations & \multicolumn{3}{c}{670} & \multicolumn{3}{c}{670} \\
\bottomrule
                \multicolumn{7}{l}{\textit{\textbf{* $p<0.05$ \quad ** $p<0.01$ \quad *** $p<0.001$}}} \\
                \end{tabular}
                \end{adjustbox}
                
\caption{Results from mixed-effects models for Male, based on 2 days of desktop usage data.}
\label{tab:mixed_effects_male for last 2 of desktop data}
                \end{table}
                
\begin{table}[H]
\centering
\begin{adjustbox}{max width=0.80\textwidth}
\begin{tabular}{l c r r c r r}

\toprule
\textbf{Predictors} & \multicolumn{3}{c}{Model 1} & \multicolumn{3}{c}{Model 2} \\
\cmidrule(lr){2-4} \cmidrule(lr){5-7}
& Estimate & CI & $p$ & Estimate & CI & $p$ \\
\midrule
Intercept & 23.48$^{***}$ & 20.66--26.30 & \textbf{$<$0.001} & 23.44$^{***}$ & 20.62--26.26 & \textbf{$<$0.001} \\
Survey wave & -0.18 & -0.38--0.03 & 0.088 & -0.14 & -0.35--0.07 & 0.187 \\
Age & -2.23$^{***}$ & -3.27-- -1.19 & \textbf{$<$0.001} & -2.16$^{***}$ & -3.22-- -1.10 & \textbf{$<$0.001} \\
Income & -1.04$^{**}$ & -1.75-- -0.34 & \textbf{0.004} & -1.03$^{**}$ & -1.75-- -0.31 & \textbf{0.005} \\
Total time spent online & -0.02 & -0.12--0.09 & 0.759 &  &  &  \\
Daytime nighttime difference & 0.01 & -0.13--0.15 & 0.873 & 0.05 & -0.11--0.20 & 0.555 \\
Time spent on entertainment &  &  &  & 0.05 & -0.11--0.21 & 0.561 \\
Time spent on adult content &  &  &  & -0.20 & -0.59--0.19 & 0.307 \\
Time spent on social media &  &  &  & 0.02 & -0.27--0.32 & 0.874 \\
Time spent on messaging &  &  &  & -0.59 & -1.44--0.27 & 0.180 \\
Time spent on games &  &  &  & -0.10 & -0.65--0.45 & 0.729 \\
Time spent on shopping &  &  &  & -0.05 & -0.51--0.41 & 0.835 \\
Time spent on productivity &  &  &  & -0.20 & -0.61--0.21 & 0.330 \\
Time spent on news &  &  &  & -0.49 & -1.13--0.15 & 0.133 \\
\midrule
\multicolumn{7}{l}{\textbf{Random Effects}} \\
$\sigma^2$ & \multicolumn{3}{c}{11.21} & \multicolumn{3}{c}{11.25} \\
$\tau_{00}$ & \multicolumn{3}{c}{41.93$_{\text{pid}}$} & \multicolumn{3}{c}{41.81$_{\text{pid}}$} \\
ICC & \multicolumn{3}{c}{0.79} & \multicolumn{3}{c}{0.79} \\
N & \multicolumn{3}{c}{231$_{\text{pid}}$} & \multicolumn{3}{c}{231$_{\text{pid}}$} \\
Observations & \multicolumn{3}{c}{537} & \multicolumn{3}{c}{537} \\
\bottomrule
                \multicolumn{7}{l}{\textit{\textbf{* $p<0.05$ \quad ** $p<0.01$ \quad *** $p<0.001$}}} \\
                \end{tabular}
                \end{adjustbox}
                
\caption{Results from mixed-effects models for Female, based on 2 days of desktop usage data.}
\label{tab:mixed_effects_female for last 2 of desktop data}
                \end{table}


\subsection{Age Differences}


\begin{table}[H]
\centering
\begin{adjustbox}{max width=0.80\textwidth}
\begin{tabular}{l c r r c r r}
\toprule
\textbf{Predictors} & \multicolumn{3}{c}{Model 1} & \multicolumn{3}{c}{Model 2} \\
\cmidrule(lr){2-4} \cmidrule(lr){5-7}
& Estimate & CI & $p$ & Estimate & CI & $p$ \\
\midrule
Intercept & 18.88$^{***}$ & 15.54--22.22 & \textbf{$<$0.001} & 19.18$^{***}$ & 15.85--22.51 & \textbf{$<$0.001} \\
Survey wave & -0.06 & -0.36--0.23 & 0.678 & -0.13 & -0.43--0.17 & 0.399 \\
Gender & -0.39 & -3.21--2.43 & 0.785 & -0.70 & -3.63--2.22 & 0.638 \\
Income & 0.29 & -0.67--1.24 & 0.558 & 0.30 & -0.67--1.27 & 0.542 \\
Total time spent online & -0.00 & -0.01--0.01 & 0.970 &  &  &  \\
Daytime nighttime difference & -0.00 & -0.03--0.03 & 0.980 & -0.01 & -0.05--0.03 & 0.591 \\
Time spent on entertainment &  &  &  & -0.01 & -0.06--0.03 & 0.550 \\
Time spent on social media &  &  &  & -0.00 & -0.03--0.03 & 0.972 \\
Time spent on messaging &  &  &  & -0.04 & -0.09--0.02 & 0.200 \\
Time spent on games &  &  &  & 0.00 & -0.05--0.05 & 0.958 \\
Time spent on shopping &  &  &  & 0.12 & -0.02--0.25 & 0.085 \\
Time spent on productivity &  &  &  & 0.04 & -0.04--0.12 & 0.352 \\
Time spent on news &  &  &  & 0.06 & -0.23--0.36 & 0.667 \\
\midrule
\multicolumn{7}{l}{\textbf{Random Effects}} \\
$\sigma^2$ & \multicolumn{3}{c}{10.12} & \multicolumn{3}{c}{10.16} \\
$\tau_{00}$ & \multicolumn{3}{c}{20.13$_{\text{pid}}$} & \multicolumn{3}{c}{20.18$_{\text{pid}}$} \\
ICC & \multicolumn{3}{c}{0.67} & \multicolumn{3}{c}{0.67} \\
N & \multicolumn{3}{c}{53$_{\text{pid}}$} & \multicolumn{3}{c}{53$_{\text{pid}}$} \\
Observations & \multicolumn{3}{c}{197} & \multicolumn{3}{c}{197} \\
\bottomrule
                \multicolumn{7}{l}{\textit{\textbf{* $p<0.05$ \quad ** $p<0.01$ \quad *** $p<0.001$}}} \\
                \end{tabular}
                \end{adjustbox}
                
\caption{Results from mixed-effects models for Age 18–30, based on 30 days of mobile usage data.}
\label{tab:mixed_effects_age_18–30 for last 30 of mobile data}
                \end{table}
                
\begin{table}[H]
\centering
\begin{adjustbox}{max width=0.80\textwidth}
\begin{tabular}{l c r r c r r}
\toprule
\textbf{Predictors} & \multicolumn{3}{c}{Model 1} & \multicolumn{3}{c}{Model 2} \\
\cmidrule(lr){2-4} \cmidrule(lr){5-7}
& Estimate & CI & $p$ & Estimate & CI & $p$ \\
\midrule
Intercept & 19.39$^{***}$ & 17.24--21.53 & \textbf{$<$0.001} & 19.54$^{***}$ & 17.39--21.69 & \textbf{$<$0.001} \\
Survey wave & -0.14$^{*}$ & -0.28-- -0.01 & \textbf{0.040} & -0.13 & -0.27--0.00 & 0.054 \\
Gender & 1.64$^{*}$ & 0.10--3.19 & \textbf{0.037} & 1.60$^{*}$ & 0.03--3.17 & \textbf{0.046} \\
Income & -1.27$^{***}$ & -1.89-- -0.66 & \textbf{$<$0.001} & -1.24$^{***}$ & -1.85-- -0.63 & \textbf{$<$0.001} \\
Total time spent online & 0.01 & -0.00--0.01 & 0.204 &  &  &  \\
Daytime nighttime difference & -0.00 & -0.02--0.02 & 0.875 & 0.00 & -0.01--0.02 & 0.631 \\
Time spent on entertainment &  &  &  & 0.02 & -0.00--0.04 & 0.075 \\
Time spent on social media &  &  &  & -0.00 & -0.02--0.02 & 0.964 \\
Time spent on messaging &  &  &  & 0.03 & -0.00--0.06 & 0.066 \\
Time spent on games &  &  &  & 0.00 & -0.01--0.02 & 0.587 \\
Time spent on shopping &  &  &  & 0.02 & -0.04--0.09 & 0.539 \\
Time spent on productivity &  &  &  & -0.08$^{**}$ & -0.14-- -0.02 & \textbf{0.007} \\
Time spent on news &  &  &  & -0.03 & -0.13--0.08 & 0.597 \\
\midrule
\multicolumn{7}{l}{\textbf{Random Effects}} \\
$\sigma^2$ & \multicolumn{3}{c}{12.02} & \multicolumn{3}{c}{11.94} \\
$\tau_{00}$ & \multicolumn{3}{c}{33.13$_{\text{pid}}$} & \multicolumn{3}{c}{33.04$_{\text{pid}}$} \\
ICC & \multicolumn{3}{c}{0.73} & \multicolumn{3}{c}{0.73} \\
N & \multicolumn{3}{c}{246$_{\text{pid}}$} & \multicolumn{3}{c}{246$_{\text{pid}}$} \\
Observations & \multicolumn{3}{c}{974} & \multicolumn{3}{c}{974} \\
\bottomrule
                \multicolumn{7}{l}{\textit{\textbf{* $p<0.05$ \quad ** $p<0.01$ \quad *** $p<0.001$}}} \\
                \end{tabular}
                \end{adjustbox}
                
\caption{Results from mixed-effects models for Age 31–45, based on 30 days of mobile usage data.}
\label{tab:mixed_effects_age_31–45 for last 30 of mobile data}
                \end{table}
                
\begin{table}[H]
\centering
\begin{adjustbox}{max width=0.80\textwidth}
\begin{tabular}{l c r r c r r}
\toprule
\textbf{Predictors} & \multicolumn{3}{c}{Model 1} & \multicolumn{3}{c}{Model 2} \\
\cmidrule(lr){2-4} \cmidrule(lr){5-7}
& Estimate & CI & $p$ & Estimate & CI & $p$ \\
\midrule
Intercept & 18.13$^{***}$ & 15.96--20.30 & \textbf{$<$0.001} & 18.05$^{***}$ & 15.87--20.24 & \textbf{$<$0.001} \\
Survey wave & -0.09 & -0.23--0.05 & 0.198 & -0.09 & -0.23--0.05 & 0.219 \\
Gender & 2.31$^{**}$ & 0.63--3.98 & \textbf{0.007} & 2.23$^{**}$ & 0.54--3.92 & \textbf{0.010} \\
Income & -1.12$^{**}$ & -1.79-- -0.45 & \textbf{0.001} & -1.05$^{**}$ & -1.73-- -0.37 & \textbf{0.002} \\
Total time spent online & -0.01 & -0.01--0.00 & 0.215 &  &  &  \\
Daytime nighttime difference & -0.00 & -0.02--0.01 & 0.551 & -0.00 & -0.02--0.01 & 0.566 \\
Time spent on entertainment &  &  &  & -0.00 & -0.03--0.03 & 0.978 \\
Time spent on social media &  &  &  & 0.00 & -0.02--0.02 & 0.774 \\
Time spent on messaging &  &  &  & -0.02 & -0.05--0.01 & 0.158 \\
Time spent on games &  &  &  & -0.00 & -0.01--0.01 & 0.965 \\
Time spent on shopping &  &  &  & 0.01 & -0.06--0.08 & 0.788 \\
Time spent on productivity &  &  &  & -0.03 & -0.08--0.02 & 0.244 \\
Time spent on news &  &  &  & -0.04 & -0.14--0.05 & 0.384 \\
\midrule
\multicolumn{7}{l}{\textbf{Random Effects}} \\
$\sigma^2$ & \multicolumn{3}{c}{12.53} & \multicolumn{3}{c}{12.55} \\
$\tau_{00}$ & \multicolumn{3}{c}{39.64$_{\text{pid}}$} & \multicolumn{3}{c}{39.97$_{\text{pid}}$} \\
ICC & \multicolumn{3}{c}{0.76} & \multicolumn{3}{c}{0.76} \\
N & \multicolumn{3}{c}{247$_{\text{pid}}$} & \multicolumn{3}{c}{247$_{\text{pid}}$} \\
Observations & \multicolumn{3}{c}{991} & \multicolumn{3}{c}{991} \\
\bottomrule
                \multicolumn{7}{l}{\textit{\textbf{* $p<0.05$ \quad ** $p<0.01$ \quad *** $p<0.001$}}} \\
                \end{tabular}
                \end{adjustbox}
                
\caption{Results from mixed-effects models for Age 46–60, based on 30 days of mobile usage data.}
\label{tab:mixed_effects_age_46–60 for last 30 of mobile data}
                \end{table}
                
\begin{table}[H]
\centering
\begin{adjustbox}{max width=0.80\textwidth}
\begin{tabular}{l c r r c r r}
\toprule
\textbf{Predictors} & \multicolumn{3}{c}{Model 1} & \multicolumn{3}{c}{Model 2} \\
\cmidrule(lr){2-4} \cmidrule(lr){5-7}
& Estimate & CI & $p$ & Estimate & CI & $p$ \\
\midrule
Intercept & 15.59$^{***}$ & 12.33--18.84 & \textbf{$<$0.001} & 15.58$^{***}$ & 12.29--18.87 & \textbf{$<$0.001} \\
Survey wave & -0.07 & -0.25--0.11 & 0.442 & -0.09 & -0.26--0.09 & 0.348 \\
Gender & 1.45 & -1.24--4.15 & 0.291 & 1.12 & -1.65--3.89 & 0.428 \\
Income & -1.44$^{**}$ & -2.49-- -0.39 & \textbf{0.007} & -1.42$^{**}$ & -2.48-- -0.35 & \textbf{0.009} \\
Total time spent online & 0.01$^{*}$ & 0.00--0.03 & \textbf{0.029} &  &  &  \\
Daytime nighttime difference & -0.03$^{*}$ & -0.05-- -0.01 & \textbf{0.012} & -0.03$^{*}$ & -0.05-- -0.01 & \textbf{0.012} \\
Time spent on entertainment &  &  &  & -0.00 & -0.06--0.06 & 0.986 \\
Time spent on social media &  &  &  & 0.01 & -0.03--0.04 & 0.722 \\
Time spent on messaging &  &  &  & 0.05 & -0.00--0.09 & 0.057 \\
Time spent on games &  &  &  & 0.02 & -0.01--0.04 & 0.169 \\
Time spent on shopping &  &  &  & 0.10$^{*}$ & 0.02--0.19 & \textbf{0.021} \\
Time spent on productivity &  &  &  & -0.01 & -0.07--0.05 & 0.769 \\
Time spent on news &  &  &  & -0.02 & -0.15--0.10 & 0.700 \\
\midrule
\multicolumn{7}{l}{\textbf{Random Effects}} \\
$\sigma^2$ & \multicolumn{3}{c}{9.67} & \multicolumn{3}{c}{9.59} \\
$\tau_{00}$ & \multicolumn{3}{c}{42.62$_{\text{pid}}$} & \multicolumn{3}{c}{43.17$_{\text{pid}}$} \\
ICC & \multicolumn{3}{c}{0.82} & \multicolumn{3}{c}{0.82} \\
N & \multicolumn{3}{c}{110$_{\text{pid}}$} & \multicolumn{3}{c}{110$_{\text{pid}}$} \\
Observations & \multicolumn{3}{c}{438} & \multicolumn{3}{c}{438} \\
\bottomrule
                \multicolumn{7}{l}{\textit{\textbf{* $p<0.05$ \quad ** $p<0.01$ \quad *** $p<0.001$}}} \\
                \end{tabular}
                \end{adjustbox}
                
\caption{Results from mixed-effects models for Age 60+, based on 30 days of mobile usage data.}
\label{tab:mixed_effects_age_60+ for last 30 of mobile data}
                \end{table}


\begin{table}[H]
\centering
\begin{adjustbox}{max width=0.80\textwidth}
\begin{tabular}{l c r r c r r}

\toprule
\textbf{Predictors} & \multicolumn{3}{c}{Model 1} & \multicolumn{3}{c}{Model 2} \\
\cmidrule(lr){2-4} \cmidrule(lr){5-7}
& Estimate & CI & $p$ & Estimate & CI & $p$ \\
\midrule
Intercept & 16.85$^{***}$ & 10.82--22.89 & \textbf{$<$0.001} & 17.39$^{***}$ & 10.99--23.79 & \textbf{$<$0.001} \\
Survey wave & -0.08 & -0.66--0.49 & 0.773 & 0.18 & -0.44--0.81 & 0.565 \\
Gender & 2.88 & -1.80--7.56 & 0.228 & 1.83 & -3.26--6.92 & 0.480 \\
Income & 0.02 & -1.51--1.55 & 0.976 & -0.12 & -1.76--1.53 & 0.891 \\
Total time spent online & 0.01 & -0.03--0.05 & 0.724 &  &  &  \\
Daytime nighttime difference & 0.06 & -0.00--0.13 & 0.063 & 0.07 & -0.00--0.14 & 0.058 \\
Time spent on entertainment &  &  &  & 0.05 & -0.05--0.16 & 0.316 \\
Time spent on adult content &  &  &  & -0.87 & -1.77--0.03 & 0.058 \\
Time spent on social media &  &  &  & -0.15 & -0.49--0.19 & 0.380 \\
Time spent on messaging &  &  &  & -0.25 & -0.83--0.33 & 0.394 \\
Time spent on games &  &  &  & 1.08 & -0.25--2.40 & 0.111 \\
Time spent on shopping &  &  &  & 0.11 & -0.34--0.56 & 0.645 \\
Time spent on productivity &  &  &  & -0.08 & -0.28--0.13 & 0.466 \\
Time spent on news &  &  &  & 1.19 & -0.65--3.04 & 0.205 \\
\midrule
\multicolumn{7}{l}{\textbf{Random Effects}} \\
$\sigma^2$ & \multicolumn{3}{c}{12.96} & \multicolumn{3}{c}{11.68} \\
$\tau_{00}$ & \multicolumn{3}{c}{33.07$_{\text{pid}}$} & \multicolumn{3}{c}{36.07$_{\text{pid}}$} \\
ICC & \multicolumn{3}{c}{0.72} & \multicolumn{3}{c}{0.76} \\
N & \multicolumn{3}{c}{38$_{\text{pid}}$} & \multicolumn{3}{c}{38$_{\text{pid}}$} \\
Observations & \multicolumn{3}{c}{85} & \multicolumn{3}{c}{85} \\
\bottomrule
                \multicolumn{7}{l}{\textit{\textbf{* $p<0.05$ \quad ** $p<0.01$ \quad *** $p<0.001$}}} \\
                \end{tabular}
                \end{adjustbox}
                
\caption{Results from mixed-effects models for Age 18–30, based on 30 days of desktop usage data.}
\label{tab:mixed_effects_age_18–30 for last 30 of desktop data}
                \end{table}
                
\begin{table}[H]
\centering
\begin{adjustbox}{max width=0.80\textwidth}
\begin{tabular}{l c r r c r r}

\toprule
\textbf{Predictors} & \multicolumn{3}{c}{Model 1} & \multicolumn{3}{c}{Model 2} \\
\cmidrule(lr){2-4} \cmidrule(lr){5-7}
& Estimate & CI & $p$ & Estimate & CI & $p$ \\
\midrule
Intercept & 19.49$^{***}$ & 17.10--21.88 & \textbf{$<$0.001} & 19.29$^{***}$ & 16.90--21.68 & \textbf{$<$0.001} \\
Survey wave & -0.42$^{***}$ & -0.65-- -0.19 & \textbf{$<$0.001} & -0.38$^{**}$ & -0.61-- -0.14 & \textbf{0.002} \\
Gender & 2.92$^{**}$ & 0.85--4.98 & \textbf{0.006} & 2.49$^{*}$ & 0.37--4.60 & \textbf{0.021} \\
Income & -0.96$^{*}$ & -1.72-- -0.20 & \textbf{0.013} & -0.87$^{*}$ & -1.63-- -0.10 & \textbf{0.027} \\
Total time spent online & -0.01 & -0.01--0.00 & 0.207 &  &  &  \\
Daytime nighttime difference & -0.01 & -0.03--0.01 & 0.549 & 0.00 & -0.02--0.02 & 0.914 \\
Time spent on entertainment &  &  &  & 0.00 & -0.01--0.01 & 0.777 \\
Time spent on adult content &  &  &  & -0.01 & -0.03--0.02 & 0.615 \\
Time spent on social media &  &  &  & 0.01 & -0.06--0.09 & 0.736 \\
Time spent on messaging &  &  &  & 0.10 & -0.14--0.34 & 0.406 \\
Time spent on games &  &  &  & 0.00 & -0.05--0.06 & 0.852 \\
Time spent on shopping &  &  &  & -0.01 & -0.07--0.06 & 0.859 \\
Time spent on productivity &  &  &  & 0.01 & -0.04--0.07 & 0.665 \\
Time spent on news &  &  &  & -0.26$^{**}$ & -0.42-- -0.10 & \textbf{0.002} \\
\midrule
\multicolumn{7}{l}{\textbf{Random Effects}} \\
$\sigma^2$ & \multicolumn{3}{c}{10.26} & \multicolumn{3}{c}{10.03} \\
$\tau_{00}$ & \multicolumn{3}{c}{31.95$_{\text{pid}}$} & \multicolumn{3}{c}{32.95$_{\text{pid}}$} \\
ICC & \multicolumn{3}{c}{0.76} & \multicolumn{3}{c}{0.77} \\
N & \multicolumn{3}{c}{137$_{\text{pid}}$} & \multicolumn{3}{c}{137$_{\text{pid}}$} \\
Observations & \multicolumn{3}{c}{395} & \multicolumn{3}{c}{395} \\
\bottomrule
                \multicolumn{7}{l}{\textit{\textbf{* $p<0.05$ \quad ** $p<0.01$ \quad *** $p<0.001$}}} \\
                \end{tabular}
                \end{adjustbox}
                
\caption{Results from mixed-effects models for Age 31–45, based on 30 days of desktop usage data.}
\label{tab:mixed_effects_age_31–45 for last 30 of desktop data}
                \end{table}
                
\begin{table}[H]
\centering
\begin{adjustbox}{max width=0.80\textwidth}
\begin{tabular}{l c r r c r r}

\toprule
\textbf{Predictors} & \multicolumn{3}{c}{Model 1} & \multicolumn{3}{c}{Model 2} \\
\cmidrule(lr){2-4} \cmidrule(lr){5-7}
& Estimate & CI & $p$ & Estimate & CI & $p$ \\
\midrule
Intercept & 17.29$^{***}$ & 15.05--19.54 & \textbf{$<$0.001} & 17.37$^{***}$ & 15.11--19.63 & \textbf{$<$0.001} \\
Survey wave & -0.08 & -0.24--0.07 & 0.281 & -0.10 & -0.26--0.05 & 0.191 \\
Gender & 1.49 & -0.42--3.41 & 0.127 & 1.28 & -0.64--3.20 & 0.192 \\
Income & -1.04$^{**}$ & -1.76-- -0.33 & \textbf{0.004} & -1.04$^{**}$ & -1.77-- -0.31 & \textbf{0.005} \\
Total time spent online & 0.00 & -0.01--0.01 & 0.824 &  &  &  \\
Daytime nighttime difference & 0.01 & -0.01--0.02 & 0.397 & 0.01 & -0.01--0.02 & 0.332 \\
Time spent on entertainment &  &  &  & -0.01 & -0.04--0.01 & 0.348 \\
Time spent on adult content &  &  &  & -0.02 & -0.05--0.01 & 0.162 \\
Time spent on social media &  &  &  & 0.01 & -0.02--0.04 & 0.448 \\
Time spent on messaging &  &  &  & 0.00 & -0.06--0.06 & 0.958 \\
Time spent on games &  &  &  & -0.02 & -0.07--0.03 & 0.385 \\
Time spent on shopping &  &  &  & 0.04 & -0.00--0.09 & 0.062 \\
Time spent on productivity &  &  &  & -0.01 & -0.05--0.03 & 0.582 \\
Time spent on news &  &  &  & 0.02 & -0.04--0.08 & 0.504 \\
\midrule
\multicolumn{7}{l}{\textbf{Random Effects}} \\
$\sigma^2$ & \multicolumn{3}{c}{11.36} & \multicolumn{3}{c}{11.36} \\
$\tau_{00}$ & \multicolumn{3}{c}{49.33$_{\text{pid}}$} & \multicolumn{3}{c}{49.04$_{\text{pid}}$} \\
ICC & \multicolumn{3}{c}{0.81} & \multicolumn{3}{c}{0.81} \\
N & \multicolumn{3}{c}{232$_{\text{pid}}$} & \multicolumn{3}{c}{232$_{\text{pid}}$} \\
Observations & \multicolumn{3}{c}{790} & \multicolumn{3}{c}{790} \\
\bottomrule
                \multicolumn{7}{l}{\textit{\textbf{* $p<0.05$ \quad ** $p<0.01$ \quad *** $p<0.001$}}} \\
                \end{tabular}
                \end{adjustbox}
                
\caption{Results from mixed-effects models for Age 46–60, based on 30 days of desktop usage data.}
\label{tab:mixed_effects_age_46–60 for last 30 of desktop data}
                \end{table}
                
\begin{table}[H]
\centering
\begin{adjustbox}{max width=0.80\textwidth}
\begin{tabular}{l c r r c r r}

\toprule
\textbf{Predictors} & \multicolumn{3}{c}{Model 1} & \multicolumn{3}{c}{Model 2} \\
\cmidrule(lr){2-4} \cmidrule(lr){5-7}
& Estimate & CI & $p$ & Estimate & CI & $p$ \\
\midrule
Intercept & 16.15$^{***}$ & 13.35--18.94 & \textbf{$<$0.001} & 16.37$^{***}$ & 13.54--19.20 & \textbf{$<$0.001} \\
Survey wave & -0.06 & -0.26--0.13 & 0.536 & -0.07 & -0.26--0.13 & 0.523 \\
Gender & 1.34 & -1.07--3.76 & 0.276 & 1.03 & -1.43--3.48 & 0.413 \\
Income & -1.31$^{**}$ & -2.24-- -0.38 & \textbf{0.006} & -1.33$^{**}$ & -2.29-- -0.37 & \textbf{0.006} \\
Total time spent online & -0.00 & -0.02--0.01 & 0.722 &  &  &  \\
Daytime nighttime difference & -0.01 & -0.03--0.01 & 0.435 & -0.01 & -0.04--0.01 & 0.264 \\
Time spent on entertainment &  &  &  & -0.04 & -0.09--0.01 & 0.144 \\
Time spent on adult content &  &  &  & -0.19 & -0.38--0.00 & 0.055 \\
Time spent on social media &  &  &  & -0.01 & -0.03--0.02 & 0.652 \\
Time spent on messaging &  &  &  & 0.07 & -0.04--0.18 & 0.220 \\
Time spent on games &  &  &  & 0.04 & -0.02--0.10 & 0.193 \\
Time spent on shopping &  &  &  & 0.03 & -0.04--0.11 & 0.404 \\
Time spent on productivity &  &  &  & 0.02 & -0.03--0.06 & 0.505 \\
Time spent on news &  &  &  & -0.04 & -0.09--0.01 & 0.150 \\
\midrule
\multicolumn{7}{l}{\textbf{Random Effects}} \\
$\sigma^2$ & \multicolumn{3}{c}{11.00} & \multicolumn{3}{c}{10.81} \\
$\tau_{00}$ & \multicolumn{3}{c}{35.97$_{\text{pid}}$} & \multicolumn{3}{c}{36.63$_{\text{pid}}$} \\
ICC & \multicolumn{3}{c}{0.77} & \multicolumn{3}{c}{0.77} \\
N & \multicolumn{3}{c}{119$_{\text{pid}}$} & \multicolumn{3}{c}{119$_{\text{pid}}$} \\
Observations & \multicolumn{3}{c}{443} & \multicolumn{3}{c}{443} \\
\bottomrule
                \multicolumn{7}{l}{\textit{\textbf{* $p<0.05$ \quad ** $p<0.01$ \quad *** $p<0.001$}}} \\
                \end{tabular}
                \end{adjustbox}
                
\caption{Results from mixed-effects models for Age 60+, based on 30 days of desktop usage data.}
\label{tab:mixed_effects_age_60+ for last 30 of desktop data}
                \end{table}


\begin{table}[H]
\centering
\begin{adjustbox}{max width=0.80\textwidth}
\begin{tabular}{l c r r c r r}
\toprule
\textbf{Predictors} & \multicolumn{3}{c}{Model 1} & \multicolumn{3}{c}{Model 2} \\
\cmidrule(lr){2-4} \cmidrule(lr){5-7}
& Estimate & CI & $p$ & Estimate & CI & $p$ \\
\midrule
Intercept & 18.96$^{***}$ & 15.32--22.61 & \textbf{$<$0.001} & 19.09$^{***}$ & 15.33--22.85 & \textbf{$<$0.001} \\
Survey wave & -0.21 & -0.56--0.14 & 0.236 & -0.24 & -0.62--0.13 & 0.205 \\
Gender & -0.16 & -3.20--2.88 & 0.919 & -0.16 & -3.29--2.98 & 0.923 \\
Income & 0.22 & -0.85--1.28 & 0.688 & 0.16 & -0.93--1.26 & 0.767 \\
Total time spent online & 0.00 & -0.12--0.13 & 0.944 &  &  &  \\
Daytime nighttime difference & 0.08 & -0.19--0.36 & 0.547 & 0.12 & -0.17--0.42 & 0.421 \\
Time spent on entertainment &  &  &  & -0.11 & -0.55--0.32 & 0.612 \\
Time spent on social media &  &  &  & -0.02 & -0.40--0.36 & 0.904 \\
Time spent on messaging &  &  &  & 0.15 & -0.40--0.69 & 0.597 \\
Time spent on games &  &  &  & 0.06 & -0.44--0.56 & 0.800 \\
Time spent on shopping &  &  &  & 1.18 & -0.43--2.79 & 0.150 \\
Time spent on productivity &  &  &  & -0.72 & -1.79--0.35 & 0.187 \\
Time spent on news &  &  &  & -0.06 & -2.69--2.56 & 0.962 \\
\midrule
\multicolumn{7}{l}{\textbf{Random Effects}} \\
$\sigma^2$ & \multicolumn{3}{c}{10.97} & \multicolumn{3}{c}{11.02} \\
$\tau_{00}$ & \multicolumn{3}{c}{20.90$_{\text{pid}}$} & \multicolumn{3}{c}{21.79$_{\text{pid}}$} \\
ICC & \multicolumn{3}{c}{0.66} & \multicolumn{3}{c}{0.66} \\
N & \multicolumn{3}{c}{48$_{\text{pid}}$} & \multicolumn{3}{c}{48$_{\text{pid}}$} \\
Observations & \multicolumn{3}{c}{163} & \multicolumn{3}{c}{163} \\
\bottomrule
                \multicolumn{7}{l}{\textit{\textbf{* $p<0.05$ \quad ** $p<0.01$ \quad *** $p<0.001$}}} \\
                \end{tabular}%
                \end{adjustbox}
                
\caption{Results from mixed-effects models for Age 60+, based on 2 days of mobile usage data.}
\label{tab:mixed_effects_age_60+ for last 2 of mobile data}
                \end{table}


\begin{table}[H]
\centering
\begin{adjustbox}{max width=0.80\textwidth}
%
                \end{adjustbox}
                
\caption{Results from mixed-effects models for Age 60+, based on 2 days of desktop usage data.}
\label{tab:mixed_effects_age_60+ for last 2 of desktop data}
                \end{table}

\subsection{Income Differences}


\begin{table}[H]
\centering
\begin{adjustbox}{max width=0.80\textwidth}
%
                \end{adjustbox}
                
\caption{Results from mixed-effects models for Tier I income group, based on 30 days of mobile usage data. }
\label{tab:Results from mixed-effects models for tier 1, based on 30 days of mobile usage data.}

                \end{table}
                
\begin{table}[H]
\centering
\begin{adjustbox}{max width=0.80\textwidth}
%
                \end{adjustbox}
                
\caption{Results from mixed-effects models for Tier II income group, based on 30 days of mobile usage data.}
\label{Results from mixed-effects models for tier 2, based on 30 days of mobile usage data.}

                \end{table}
                
\begin{table}[H]
\centering
\begin{adjustbox}{max width=0.80\textwidth}
%
                \end{adjustbox}
                
\caption{Results from mixed-effects models for Tier III income group, based on 30 days of mobile usage data.}
\label{Results from mixed-effects models for tier 3, based on 30 days of mobile usage data.}

                \end{table}
                
\begin{table}[H]
\centering
\begin{adjustbox}{max width=0.80\textwidth}
%
                \end{adjustbox}
                
\caption{Results from mixed-effects models for Tier IV income group, based on 30 days of mobile usage data.}
\label{Results from mixed-effects models for tier 4, based on 30 days of mobile usage data.}

                \end{table}
                
\begin{table}[H]
\centering
\begin{adjustbox}{max width=0.80\textwidth}
%
                \end{adjustbox}
                
\caption{Results from mixed-effects models for Tier V income group, based on 30 days of mobile usage data.}
\label{tab:Results from mixed-effects models for tier 5, based on 30 days of mobile usage data.}

                \end{table}


\begin{table}[H]
\centering
\begin{adjustbox}{max width=0.80\textwidth}
%
                \end{adjustbox}
                
\caption{Results from mixed-effects models for Tier I, based on 30 days of desktop usage data.}
\label{tab:Results from mixed-effects models for tier 1, based on 30 days of desktop usage data.}
                \end{table}
                
\begin{table}[H]
\centering
\begin{adjustbox}{max width=0.80\textwidth}
%
                \end{adjustbox}
                
\caption{Results from mixed-effects models for Tier II, based on 30 days of desktop usage data.}
\label{Results from mixed-effects models for tier 2, based on 30 days of desktop usage data.}

                \end{table}
                
\begin{table}[H]
\centering
\begin{adjustbox}{max width=0.80\textwidth}
%
                \end{adjustbox}
                
\caption{Results from mixed-effects models for Tier III, based on 30 days of desktop usage data.}
\label{Results from mixed-effects models for tier 3, based on 30 days of desktop usage data.}

                \end{table}
                
\begin{table}[H]
\centering
\begin{adjustbox}{max width=0.80\textwidth}
%
                \end{adjustbox}
                
\caption{Results from mixed-effects models for Tier IV, based on 30 days of desktop usage data.}
\label{Results from mixed-effects models for tier 4, based on 30 days of desktop usage data.}

                \end{table}
                
\begin{table}[H]
\centering
\begin{adjustbox}{max width=0.80\textwidth}
%
                \end{adjustbox}
                
\caption{Results from mixed-effects models for Tier V, based on 30 days of desktop usage data.}
\label{tab:Results from mixed-effects models for tier 5, based on 30 days of desktop usage data.}

                \end{table}


\begin{table}[H]
\centering
\begin{adjustbox}{max width=0.80\textwidth}
%
                \end{adjustbox}
                
\caption{Results from mixed-effects models for Tier I income group, based on 2 days of mobile usage data.}
\label{tab:Results from mixed-effects models for tier 1, based on 2 days of mobile usage data.}

                \end{table}
                
\begin{table}[H]
\centering
\begin{adjustbox}{max width=0.80\textwidth}
%
                \end{adjustbox}
                
\caption{Results from mixed-effects models for Tier II income group, based on 2 days of mobile usage data.}
\label{Results from mixed-effects models for tier 2, based on 2 days of mobile usage data.}

                \end{table}
                
\begin{table}[H]
\centering
\begin{adjustbox}{max width=0.80\textwidth}
%
                \end{adjustbox}
                
\caption{Results from mixed-effects models for Tier III income group, based on 2 days of mobile usage data.}
\label{Results from mixed-effects models for tier 3, based on 2 days of mobile usage data.}

                \end{table}
                
\begin{table}[H]
\centering
\begin{adjustbox}{max width=0.80\textwidth}
%
                \end{adjustbox}
                
\caption{Results from mixed-effects models for Tier IV income group, based on 2 days of mobile usage data.}
\label{Results from mixed-effects models for tier 4, based on 2 days of mobile usage data.}

                \end{table}
                
\begin{table}[H]
\centering
\begin{adjustbox}{max width=0.80\textwidth}
%
                \end{adjustbox}
                
\caption{Results from mixed-effects models for Tier V income group, based on 2 days of mobile usage data.}
\label{tab:mixed_effects_Tier V for last 2 of mobile data}

\end{table}


\begin{table}[H]
\centering
\begin{adjustbox}{max width=0.80\textwidth}
%
                \end{adjustbox}
                
\caption{Results from mixed-effects models for Tier I, based on 2 days of desktop usage data.}
\label{tab:Results from mixed-effects models for tier 1, based on 2 days of desktop usage data.}

                \end{table}
                
\begin{table}[H]  
\centering
\begin{adjustbox}{max width=0.80\textwidth}
%
                \end{adjustbox}
                
\caption{Results from mixed-effects models for Tier II, based on 2 days of desktop usage data.}
\label{Results from mixed-effects models for tier 2, based on 2 days of desktop usage data.}

\end{table}
                
\begin{table}[H]
\centering
\begin{adjustbox}{max width=0.80\textwidth}
%
                \end{adjustbox}
                
\caption{Results from mixed-effects models for Tier III, based on 2 days of desktop usage data.}
\label{Results from mixed-effects models for tier 3, based on 2 days of desktop usage data.}

                \end{table}
                
\begin{table}[H]
\centering
\begin{adjustbox}{max width=0.80\textwidth}
%
                \end{adjustbox}
                
\caption{Results from mixed-effects models for Tier IV, based on 2 days of desktop usage data.}
\label{Results from mixed-effects models for tier 4, based on 2 days of desktop usage data.}

                \end{table}
                
\begin{table}[H]
\centering
\begin{adjustbox}{max width=0.80\textwidth}
%
                \end{adjustbox}
                
\caption{Results from mixed-effects models for Tier V, based on 2 days of desktop usage data.}
\label{tab:Results from mixed-effects models for tier 5, based on 2 days of desktop usage data.}

                \end{table}



\end{document}